\documentclass[aps,prd,superscriptaddress,showpacs,nofootinbib,twocolumn,floatfix,preprintnumbers,]{revtex4-1}

\usepackage{mciteplus}
\usepackage{latexsym}
\usepackage{amsthm}
\usepackage{amsmath}
\usepackage{amssymb}
\usepackage{hepunits}
\usepackage{hyperref}
\usepackage{bbm}
\usepackage{bm}
\usepackage{xfrac}
\usepackage{color}
\usepackage{colordvi}
\usepackage{comment}
\usepackage{dcolumn}
\usepackage{times,latexsym,graphicx,wrapfig,url}
\usepackage{epsfig,lineno,bm}
\usepackage{subfigure}
\usepackage{slashed}

\def\epm{e^+e^-}
\def\ap{A^\prime}

\newcommand{\s}{{\rm s}}

\newcommand{\be}{\begin{eqnarray}}
\newcommand{\ee}{\end{eqnarray}}
\newcommand{\bea}{\begin{eqnarray}}
\newcommand{\eea}{\end{eqnarray}}
\newcommand{\f}[2]{\ensuremath{\frac{#1}{#2}}}
\newcommand{\bef}{\begin{figure}[htbp]\begin{center}}
\newcommand{\eef}{\end{center}\end{figure}}

\def\lsim{\mathrel{\rlap{\lower4pt\hbox{\hskip1pt$\sim$}}
    \raise1pt\hbox{$<$}}}
\def\gsim{\mathrel{\rlap{\lower4pt\hbox{\hskip1pt$\sim$}}
    \raise1pt\hbox{$>$}}}

\newcommand\BNL{Brookhaven National Laboratory, Upton, NY }
\newcommand\Princeton{ Princeton University, Princeton, NJ }
\newcommand\FNAL{Fermi National Accelerator Laboratory, Batavia, IL }

\newcommand{\Sec}[1]{Sec.~\ref{#1}}

\newcommand{\App}[1]{App.~\ref{#1}}

\newcommand{\Eq}[1]{Eq.~(\ref{#1})}

\def\lsim{\mathrel{\rlap{\lower4pt\hbox{\hskip 0.5 pt$\sim$}}
\raise1pt\hbox{$<$}}}  

\newcommand{\apr}{A^\prime}

\newcommand{\pizero}{\pi^0}
\newcommand{\mzero}{m^0}

\begin{document}

\title{ Testing  Light Dark Matter  Coannihilation With Fixed-Target Experiments}
\preprint{FERMILAB-PUB-17-068-PPD, PUPT 2520}

\author{Eder Izaguirre}
\email{eder@bnl.gov}
\affiliation{\BNL}
\author{Yonatan Kahn}
\email{ykahn@princeton.edu}
\affiliation{\Princeton}
\author{Gordan Krnjaic}
\email{krnjaicg@fnal.gov}
\affiliation{\FNAL}
\author{Matthew Moschella}
\email{moschella@princeton.edu}
\affiliation{\Princeton}

\begin{abstract}

In this paper, we introduce a novel program of fixed-target  searches for thermal-origin Dark Matter (DM), which couples 
inelastically to the Standard Model. Since the DM only interacts by transitioning  to a heavier state, 
freeze-out proceeds via coannihilation and the unstable heavier state is depleted at later times. For sufficiently large mass splittings, 
direct detection is kinematically forbidden and indirect detection is impossible, so this scenario can only be tested with accelerators. 
Here we propose new searches at proton and electron beam fixed-target experiments to probe sub-GeV coannihilation, exploiting the distinctive signals of up- and down-scattering as well as decay of the excited state inside the detector volume. We focus on a representative model in which DM is a pseudo-Dirac fermion coupled to a hidden gauge field (dark photon), which kinetically mixes with the  visible photon. We define theoretical targets in this framework and determine the existing bounds by reanalyzing results from previous experiments. We find that LSND, E137, and BaBar data already place strong constraints on the parameter space consistent with a thermal freeze-out origin, and that future searches at Belle II and MiniBooNE, as well as recently-proposed fixed-target experiments such as LDMX and BDX, can cover nearly all remaining gaps. 
We also briefly comment on the discovery potential for proposed beam dump and neutrino experiments which operate at much higher beam energies.
\end{abstract}

\maketitle

\section{Introduction}
\label{sec:Introduction}

Understanding the particle nature of Dark Matter (DM) is among the highest priorities in all of physics. Perhaps the most 
popular DM candidate to date has been the Weakly Interacting Massive Particle (WIMP), which is charged under the 
electroweak force and naturally yields the observed cosmological abundance via thermal freeze-out (see \cite{Jungman:1995df} for a review).  However,  decades of null results from direct detection, indirect detection, and collider searches have cast doubt on this 
paradigm and motivated many alternative possibilities \cite{Alexander:2016aln,Essig:2013lka}. 

Nonetheless, the thermal freeze-out mechanism remains compelling even if DM is not a WIMP.
First and foremost, thermal DM is largely UV insensitive; its abundance is
determined by the DM particle properties and 
is unaffected by the details of earlier, unknown cosmological epochs (e.g. reheating).
Furthermore, unlike nonthermal production mechanisms, which can accommodate DM masses 
anywhere between $10^{-22}\, \eV - 10\, M_\odot$ (!),
thermal DM is only viable between $\sim 5 ~\keV  -100 ~ \TeV$, and is therefore more predictive. Dark matter masses outside this window are either too hot for acceptable structure formation \cite{Irsic:2017ixq}
or violate perturbative unitarity \cite{Griest:1989wd}.
Finally, achieving the observed abundance requires a minimum interaction rate between DM and the Standard Model (SM), which
 provides a clear target for discovery or falsification.
Thus, there is ample motivation to identify and study every viable realization of this 
mechanism. 

One simple way to completely eliminate the tension between a thermal origin and experimental limits, in particular those from direct detection experiments, is 
for the DM to couple {\it inelastically} to SM particles \cite{TuckerSmith:2001hy}. In this class of models, the halo DM species 
$\chi_1$ is the lightest stable particle in the dark sector and interacts with the SM only by
transitioning to a slightly heavier state $\chi_2$. This class of models has several appealing features:

\begin{itemize} 

\item{\bf Large Viable Couplings:}  If the inelastic interaction with the SM
also determines the leading annihilation process, the relic abundance arises dominantly through $\chi_1 \chi_2 \to {\rm \small SM}$, a process dubbed 
\emph{coannihilation} \cite{Griest:1990kh}. Since the heavier $\chi_2$ population is Boltzmann-suppressed during freeze-out, the requisite 
annihilation rate must compensate for this penalty.\footnote{For a general scenario where coannihilation proceeds without inelastic couplings, but an analogous enhanced thermal cross section appears, see Ref.~\cite{DAgnolo}.} Thus, the coannihilation cross section always satisfies $\sigma v \ge 3 \times 10^{-26} {\rm cm}^3 \s^{-1}$.

\item{\bf Indirect Detection Shuts Off:} Since the heavier state is unstable, its
population is fully depleted at low temperatures, so there are no remaining coannihilation partners for the $\chi_1$. 
This effect turns off possible late-time indirect detection signals and alleviates the bound from cosmic microwave background (CMB) power injection, which otherwise naively rules out thermal DM for masses below $\sim 10$ GeV for  $s$-wave  annihilation \cite{Ade:2015xua}.\footnote{Another way to evade this bound is ``forbidden DM'' \cite{DAgnolo:2015ujb}.}

\item{\bf Direct Detection Forbidden:} 
For a nonrelativistic halo particle scattering off a stationary SM target, the energy available to upscatter into the heavier state is $\sim \mu v^2 $, where 
$\mu$ is the reduced mass of the DM-target system.  Thus, for typical halo velocities $v\sim 10^{-3}$, direct
detection off of nuclei is forbidden if the mass splitting exceeds $ \gtrsim 100~\keV$ \cite{TuckerSmith:2001hy}. There is the possibility of loop-level induced elastic scattering off of electrons, which may be very relevant for new proposals for electron direct detection \cite{Essig:2012yx,Lee:2015qva,Essig:2015cda,Hochberg:2016ntt,Emken:2017erx,Essig:2017kqs}. However, we leave this question for a future study.

\end{itemize}

 \noindent Since direct and indirect detection search strategies are not available, testing thermal coannihilation fundamentally requires accelerator-based techniques. For DM masses near the weak scale (few GeV -- 100s of GeV), Refs. \cite{Bai:2011jg, Weiner:2012cb, Izaguirre:2015zva} proposed LHC and B-factory searches with sensitivity to thermal coannihilation over a wide range of masses and splittings.
 However, for DM masses below the GeV scale, these searches become ineffective and new tools are required to fully test the keV--GeV mass range over which thermal coannihilation remains viable, yet unexplored.  

\begin{figure*}[t] 
\hspace{0.5cm}
\includegraphics[width=13.5cm]{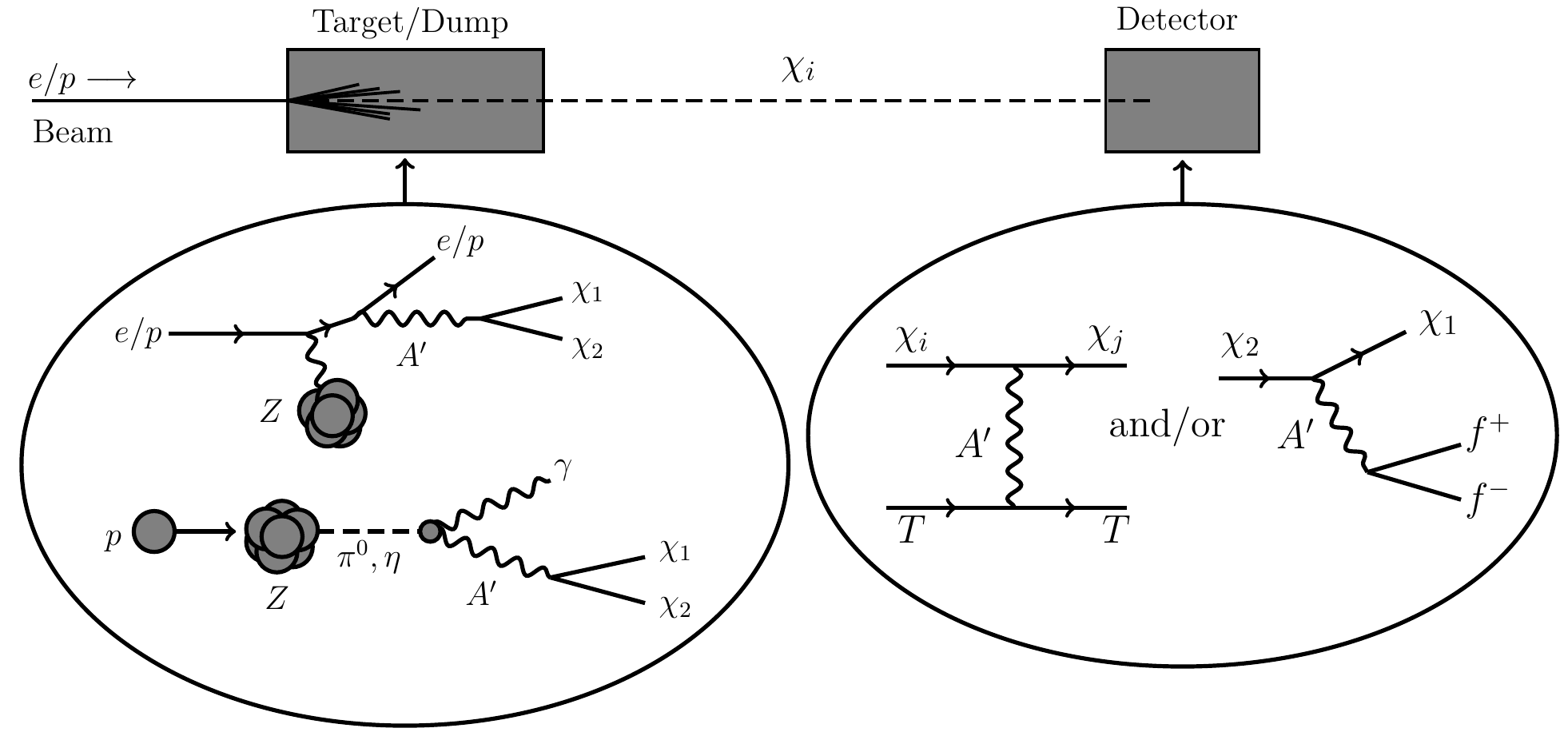}  ~~
  \caption{Inelastic DM production at electron and proton beam dump experiments via dark bremsstrahlung and meson decay. The
  resulting  $\chi_1, \chi_2$ pair can give rise to a number of 
 possible signatures in the detector: $\chi_2$ can decay inside the fiducial volume to deposit electromagnetic energy;
both $\chi_1$ and $\chi_2$ can scatter off detector targets $T$ and impart visible 
recoil energies to these particles; or $\chi_1$ can upscatter into $\chi_2$, which  can then decay promptly inside the detector to deposit a visible signal.  }
   \label{fig:beam-dump-cartoon}
\vspace{0cm}
\end{figure*}

\begin{figure*}[t] 
\hspace{0.5cm}
\includegraphics[width=14cm]{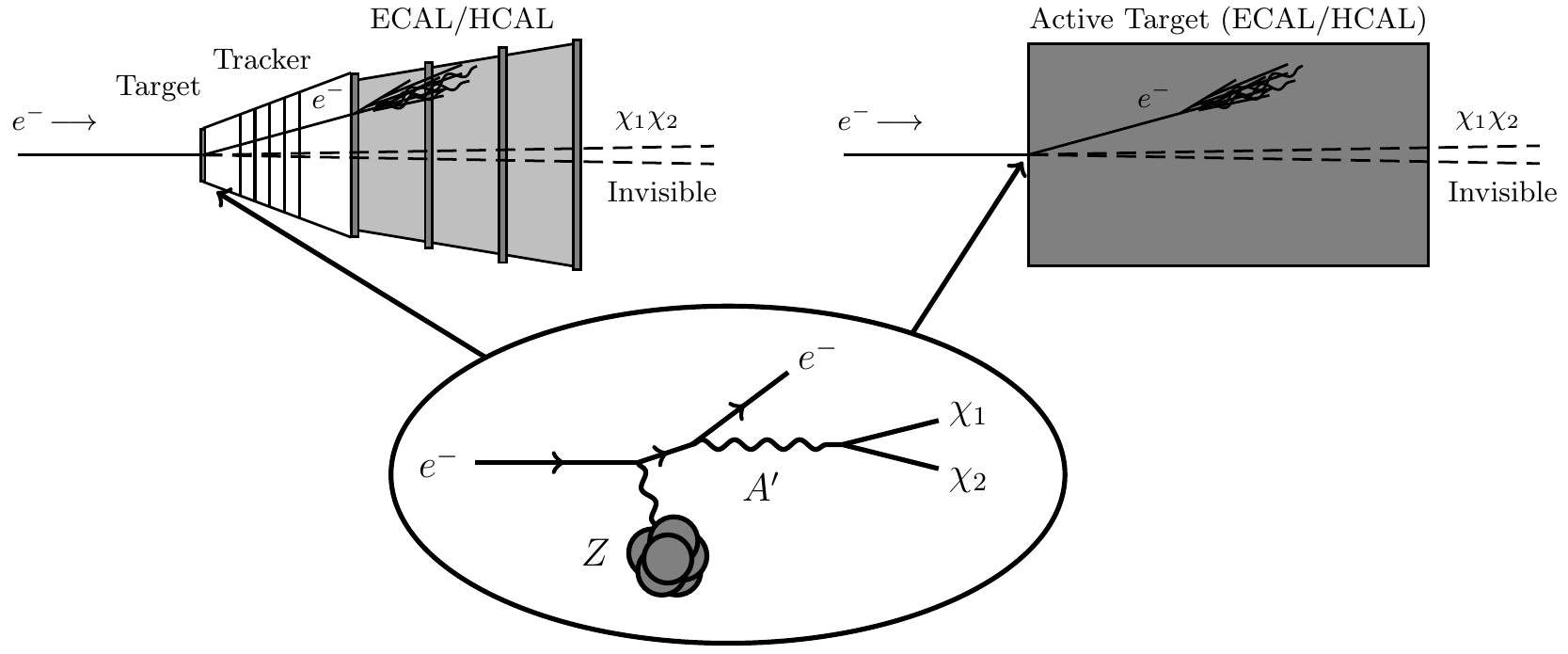}  ~~
  \caption{ Inelastic DM production at electron beam fixed-target missing energy/momentum experiments. {\bf Left:} Setup
  for an LDMX style missing momentum experiment \cite{Izaguirre:2014bca,Alexander:2016aln} in which a ($\sim$ few GeV) beam electron  
 produces DM in a thin target ($\ll$ radiation length) and thereby loses a large fraction of its incident energy. The emerging lower energy electron
 passes through tracker material and registers as a signal event if there is no additional energy deposited in the ECAL/HCAL system downstream, which 
 serves primarily to veto SM activity. {\bf Right:} Setup for an NA64 style
 experiment in which the beam (typically at higher energies, $\sim 30$ GeV) produces the DM system 
 by interacting with an instrumented, active target volume \cite{Banerjee:2016tad}. As with LDMX, the instrumented region  serves to verify that the
  beam electron has abruptly lost most of its energy and that there is no additional SM activity downstream.
   }
   \label{fig:mme-cartoon}
\vspace{0cm}
\end{figure*}

In this paper, we fill this large gap by recasting and proposing a series of fixed-target searches for both electron and proton beam facilities.  In both cases,
an incident beam impinges on a fixed target and produces a boosted $\chi_1 \chi_2$ pair.  Depending on the experimental setup, this system can give rise to a variety of possible signals:
\begin{itemize}
\item {\bf Beam Dumps:} The $\chi_1 \chi_2$ pair can be produced radiatively through the ``dark bremsstrahlung'' reactions $p N \to  p N \chi_1 \chi_2 $ at proton beam dumps, or $e N \to  e N \chi_1 \chi_2 $ at electron beam dumps, where $N$ is a nuclear target. 
At proton beam dumps, the DM system can also be produced from meson decays $\pi^0 \to \gamma \chi_1 \chi_2$ and $\eta\to \gamma \chi_1\chi_2$, which can be advantageous for low DM masses.  Once produced, the $\chi_{1,2}$ can reach a downstream detector and scatter off electrons or nucleons to induce an observable recoil signature. Alternatively, the unstable $\chi_2$ can also decay in flight as it passes through
the detector via $\chi_2 \to \chi_1e^+e^-$, thereby depositing an observable signal. These processes are depicted schematically in Fig.~\ref{fig:beam-dump-cartoon}.

\item {\bf Electron Missing Energy/Momentum:} As above,  the $\chi_1 \chi_2$ pair is produced in $e^- N \to  e^- N \chi_1 \chi_2 $ ``dark bremsstrahlung'', but the production takes place in a thin target embedded in a detector. The principal observable
signature in this context is the recoiling final-state electron. If this electron emerges having lost most of its incident beam energy, with no additional activity 
deposited in a downstream detector, this process can be sensitive to DM production. As above, a $\chi_2$ decaying inside the detector provides an additional potential signature. This process is depicted schematically in Fig.~\ref{fig:mme-cartoon}.
\end{itemize}

\begin{figure}[t] \hspace{1cm}\vspace{-0.25cm}
\includegraphics[width=5.9cm]{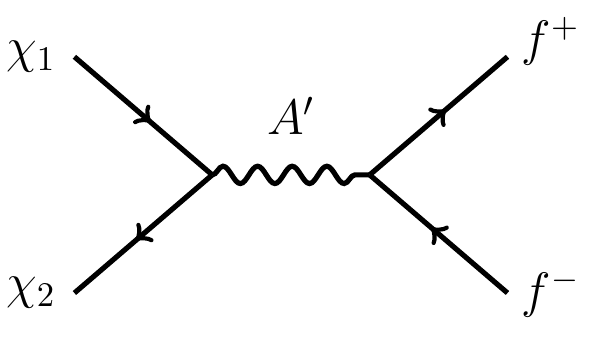}
\caption{ Leading order diagram for $\chi_1\chi_2 \to f^+f^-$ coannihilation, which
sets the DM relic abundance in the $m_{\ap} > m_{1,2}$ regime.}
\label{fig:annihilation-cartoon}
\vspace{0cm}
\end{figure}

\begin{figure}[t] 
\includegraphics[width=8.7 cm]{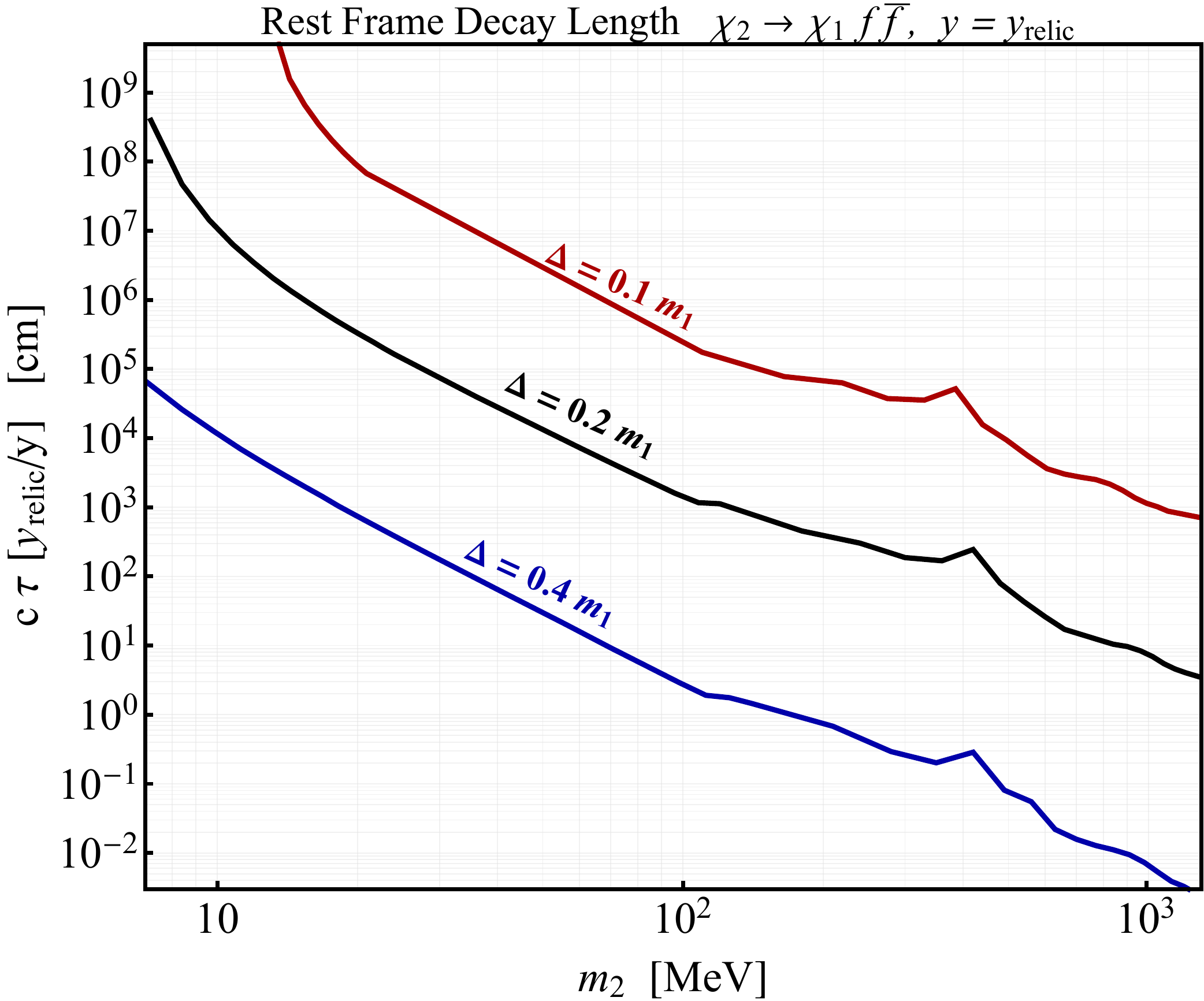}  ~~
  \caption{ Rest frame $\chi_2$ decay length for various mass splittings $\Delta\equiv m_2 - m_1$, as a function of the $\chi_2$ mass $m_2$. The vertical axis is normalized
  to values of the dimensionless coupling $y\equiv \epsilon^2 \alpha_D (\frac{m_1}{m_{\apr}})^4$ which achieve the observed relic density; see Sec.~\ref{sec:models} for more details.}   \label{fig:decaylength}
\vspace{0cm}
\end{figure}

We will show that existing data already rules out large portions of the direct coannihilation parameter space. Moreover, the dedicated searches we propose which exploit the unique signals from this class of models can significantly improve on the current levels of sensitivity. Crucially, as shown in Fig.~\ref{fig:decaylength}, the $\chi_2$ has a macroscopic decay length over nearly all of the parameter space we are interested in, and detecting the electromagnetic energy deposited by a $\chi_2$ decay in the detector --- a signal not present in elastic DM models --- is sufficient to cover large portions of the thermal relic curve.\footnote{A similar search strategy was proposed for the decay of long-lived scalars in Ref.~\cite{Schuster:2009au}.} Indeed, we show in Sec.~\ref{sec:generic} that the sensitivity to the decay of the exited state $\chi_2$ typically dominates over scattering channels at experiments with beam energies below 10 GeV. 
 Thus, the prospects are excellent for dedicated experiments sensitive to these signatures, and we show that current and planned experiments can confirm or rule out nearly the entire mass range allowed for thermal coannihilating DM. We note a related study from Ref.~\cite{Morrissey:2014yma} that investigated previous limits from fixed target facilities in the context of a supersymmetric hidden sector. Recently, Ref.~\cite{Kim:2016zjx} proposed the signal of $\chi_1 \to \chi_2$ upscattering followed by $\chi_2$ decay for the case of (boosted) astrophysical DM, as opposed to DM produced in a beam dump.

This paper is organized as follows. In Sec.~\ref{sec:models} we introduce a representative 
renormalizable model featuring a pseudo-Dirac
fermion DM coupled to a kinetically-mixed dark photon, where the abundance of the former arises from thermal coannihilation. In Sec.~\ref{sec:generic-formalism} we make some general comments on the production modes and detection signals at proton- and electron-beam fixed-target experiments. In Sec.~\ref{sec:bounds-projections} we describe our methods for reinterpreting existing data from LSND \cite{Auerbach:2001wg} and E137 \cite{Bjorken:1988as}, and we discuss projections for the future data at MiniBooNE \cite{Stancu:2001dr} and the proposed BDX \cite{Battaglieri:2014qoa} and LDMX \cite{Izaguirre:2014bca} experiments. In Sec.~\ref{sec:reach} we discuss the bounds and reach projections from these experiments in the context of the thermal coannihilating inelastic DM. Finally, in Sec.~\ref{sec:conclusion} we offer some concluding remarks. Additional details on the matrix elements and Monte Carlo methods used in determining the thermal parameter space and in our simulations can be found in the Appendices.

\section{Sub-GeV Thermal Coannihilation} 
\label{sec:models}
In this section, we describe a class of models of coannihilating DM: DM that couples inelastically to the SM through a kinetically-mixed dark photon. We detail the early universe cosmology and freeze out of the model, as well as introduce a useful parametrization of the parameters of the model in which the thermal target is largely an invariant under variation of couplings and of mass hierarchies.

\subsection{Mediator Model Building}
\label{sec:gen}

Unlike weak-scale WIMPs, which realize successful freeze-out with only SM gauge interactions, 
sub-GeV DM is overproduced in the absence of light ($\ll m_Z$) new mediators to generate a sufficiently 
large annihilation rate \cite{Lee:1977ua,Boehm:2003hm}. To avoid detection thus far, such mediators must be neutral under the SM and couple non-negligibly to visible particles.

If SM particles are neutral under the new interaction, a renormalizable model (without additional fields) requires the mediator    
to interact with the SM through the hypercharge, Higgs, or lepton portals 
\be
B_{\mu \nu} ~~,~~ H^\dagger H ~~~,   ~~~   LH, 
\ee
for vector, scalar, and fermionic mediators, respectively. 
However, coupling a fermionic mediator to the lepton portal requires additional model building\footnote{A fermionic mediator coupled to the 
lepton portal 
requires additional model building to simultaneously achieve a thermal contact through this interaction and yield viable neutrino textures; the 
coupling to the mediator must be suppressed by neutrino masses, so it is generically difficult for the interaction rate to exceed Hubble expansion. } and  
scalar mediators, which mix with the Higgs are ruled out for predictive models in which DM annihilates directly to 
SM final states (see Sec. \ref{sec:direct_vs_secluded} and \cite{Krnjaic:2015mbs} for a discussion of this issue), 
so we restrict our attention to abelian vector mediators; a nonabelian field strength is not gauge invariant, so kinetic mixing is forbidden.

Alternatively, the mediator could couple directly to SM particles if both dark and visible matter are charged under the same gauge group. 
In the absence of additional fields, anomaly cancellation restricts the possible choices to be
\be
U(1)_{B-L}~,~U(1)_{\ell_i - \ell_j}~,~  U(1)_{3B- \ell_i},
\ee 
and linear combinations thereof. In most contexts, the relevant phenomenology in fixed-target searches is qualitatively similar to
 the vector portal scenario, so below we will ignore these possibilities without loss of  essential  generality. We note, however, that viable models for both protophobic \cite{Feng:2016jff} and protophilic \cite{Tulin:2014tya} mediators exist, so the complementarity provided by both proton- and electron-beam experiments is highly advantageous.

\subsection{Representative Model}
Our representative dark sector contains a 4-component fermion $\psi$ that transforms under a hidden abelian $U(1)_D$ gauge group. The fermion
couples to a vector mediator $\ap$ as 
\be \label{eq:lag-fermion}
{\cal L} =  i \overline \psi \slashed{D} \psi + M \bar \psi \psi + \lambda \phi \, \overline{\psi^c} \psi   + h.c. ,~
\ee
where $\phi$ is a $U(1)_D$ symmetry breaking scalar whose vacuum expectation value $v_D$ gives $\apr$ a nonzero mass $m_{\apr } \sim g_D v_D$ and gives $\psi$ a Majorana mass $\sim \lambda v_D$. Diagonalizing the resulting Dirac and Majorana masses gives rise to fermion mass eigenstates  $\chi_{1,2}$ with a small mass splitting $\Delta \equiv m_2 - m_1$ and an off-diagonal coupling to $\ap$,
\be
{\cal L} \supset  g_D \ap_\mu \overline \chi_2 \gamma^\mu \chi_1  + {h.c.},
\ee
where $g_D \equiv \sqrt{4\pi \alpha_D}$ is the dark coupling constant. Note that it is technically natural to have $\Delta \ll M$ since the Majorana mass breaks the  global symmetry associated with $\psi$ number.\footnote{If, unlike the construction in Eq.~(\ref{eq:lag-fermion}), the Majorana masses for the two Weyl components in $\psi = (\xi, \eta^\dagger)$ are different, there is also a subleading diagonal interaction of the form $(\delta/M_D)  \overline \chi_i \displaystyle{\not}{\ap} \chi_i$, where $\delta \equiv m_\xi -m_\eta$ is the difference of
Majorana masses for the the interaction eigenstates. We neglect this interaction in our analysis, assuming the off-diagonal interaction dominates.}

This sector can interact with the SM through a renormalizable and gauge-invariant kinetic mixing term between $U(1)_D$ and $U(1)_Y$ gauge fields,
\be \label{eq:kinetic-mixing}
{\cal L} \supset \frac{\epsilon}{2 \cos\theta_W} F^\prime_{\mu\nu} B^{\mu \nu } = 
\epsilon F^\prime_{\mu\nu} F^{\mu \nu}  + \epsilon \tan\theta_W F^\prime_{\mu\nu} Z^{\mu \nu },~~~~
\ee
where  $\epsilon \ll 1$ is the kinetic mixing parameter and $F^\prime_{\mu \nu}$ and $B_{\mu \nu}$ are respectively the dark and hypercharge field strength tensors and the kinetic mixing interaction has been written in terms of the SM mass eigenstates $A$ and $Z$ after electroweak symmetry breaking. Diagonalizing the kinetic terms in Eq.~(\ref{eq:kinetic-mixing}) and rescaling the field strengths to restore
canonical normalization induce a coupling between $\apr$ and the SM fermions \cite{Holdom:1985ag}. To leading order in $\epsilon$, the 
$\apr$-SM interaction becomes 
\be
 eA_\mu J^\mu_{EM} \to e( A_\mu  +  \epsilon \apr_\mu )J^\mu_{EM} ,
\ee
where $J^\mu_{EM}$ is the SM electromagnetic current and all charged fermions acquire millicharges under $U(1)_D$. There is also an analogous $\apr$ interaction with the SM neutral current that arises from
$\apr-Z$ mixing, but in our mass range of interest, $m_{\apr} \ll m_Z$, the mixing parameter is suppressed by an additional 
factor of $(m_{\apr}/m_Z)^2$ \cite{Davoudiasl:2012ag, Davoudiasl:2012qa, Davoudiasl:2014kua, Kahn:2016vjr}, so we neglect this interaction for the remainder of paper.

\subsection{Direct Coannihilation vs. Secluded Annihilation}
\label{sec:direct_vs_secluded}
In the hot early universe ($T \gg m_{i}, m_{\apr}$), 
all dark species are in chemical and kinetic equilibrium with the SM radiation bath; this initial condition is guaranteed as long as the DM-SM scattering rate exceeds 
the Hubble expansion rate at some point during cosmic history. If $m_{i} > m_{\apr}$, the freeze-out process is analogous to that of WIMP models. Below the
 freeze-out temperature $T_f \sim m_{1,2}/20$, the number densities of both  species are depleted predominantly through $\chi_i \chi_i \to \apr \apr$ self-annihilation (which depends only on $\alpha_D$), \emph{not} coannihilation, which depends on the combination $\epsilon^2 \alpha_D$ and is greatly suppressed by comparison. Although both components undergo freeze-out separately, since $\chi_2$ is heavier and unstable, it will be depleted through downscattering and decays. Thus, up to order-one 
 corrections, the requisite self-annihilation cross section satisfies the familiar WIMP-like requirement $\langle \sigma v \rangle  \sim 3 \times 10^{-26} \cm^3 \, \s^{-1}$ in order
 for $\chi_1$ to have the observed abundance at late times.
 
 However, this \emph{secluded} ($m_{i} > m_{\apr}$) regime has several drawbacks. Since the self-annihilation rate for fermions is $s$-wave, annihilation 
 continues to occur out of equilibrium during recombination, which ionizes newly-formed hydrogen and thereby modifies the CMB power spectrum. 
For a thermal annihilation rate, this bound rules out DM below $\sim 10 ~\GeV$  \cite{Ade:2015xua}.\footnote{If instead, the DM is a scalar and annihilates directly to SM fermions through an $s$-channel vector mediator, its annihilation rate is $p$-wave suppressed, which can evade CMB bounds.} Furthermore, since the secluded annihilation cross section scales as $\sigma v \sim \alpha_D^2/m_{i}^2$, the abundance is independent of the $\apr$ coupling to SM states, so there is no minimum interaction strength target for the DM search effort --- at direct detection or accelerators, since these are sensitive to $\epsilon$.
 
By contrast, in the \emph{direct coannihilation} regime $(m_{i} < m_{\apr})$, the two species annihilate each other via $\chi_1 \chi_2 \to  f \overline f$, which
has several compelling features: 

\begin{itemize}
\item {\bf Predictive:} The annihilation rate depends crucially on the mixing with the SM, $\sigma v \propto \epsilon^2 \alpha_D $, so for dark couplings
 that satisfy perturbative unitarity, there is a minimum value of $\epsilon$ that is compatible with thermal freeze-out. Reaching experimental
 sensitivity to this minimum value for each viable DM mass suffices to discover or falsify this entire class of models.

\item {\bf Large Cross Section:} Unlike secluded annihilation, which involves the annihilation of equal mass particles,
direct coannihilation requires both $\chi_1$ and $\chi_2$ in the initial state. However, $\chi_2$ 
typically becomes Boltzmann-suppressed before the nominal $\chi_1$ freeze-out temperature $\sim m_1/20$, so the coannihilation cross section
must be  larger to compensate for this reduction. Thus, we generically require $\langle \sigma v \rangle  \gg 3 \times 10^{-26} \cm^3 \s^{-1}$ to achieve
the observed abundance, where the precise value grows sensitively with increasing $\Delta$ (compare the relic density curves in Fig. \ref{fig:money}).
\item{\bf Bounded Viable Range:} Since the necessary coannihilation rate for freeze-out grows sharply
 as the mass splitting is increased, for sufficiently large $\Delta$, {\it i.e.,} $\Delta \sim m_1$, the requisite $\epsilon$ is easily 
 excluded independently of other model properties (e.g.\ by precision QED/electroweak measurements). Thus, as we will see below, this class of coannihilating models can be tested over the full viable parameter space.
\end{itemize}

\noindent Thus, for the remainder of this paper, our focus will be on the direct coannihilation scenario. For simplicity, without loss of essential generality,
we will also demand  that $m_{\apr} > m_1 + m_2$ so that $\apr$ can decay to dark sector states --- otherwise, the $\apr$ decays to SM final state, a signature for which there are abundant ideas to cover \cite{Alexander:2016aln}. Since $\epsilon \ll1$, the $\apr$ branching
ratio to SM states is correspondingly negligible in this regime. 

 Note that if we had chosen a neutral scalar mediator instead of the vector $\apr$, it could only couple to SM fermions by mixing with the Higgs after
 electroweak symmetry breaking. However, such a scenario cannot viably realize thermal coannihilation below the GeV scale because the requisite Higgs-mediator
 mixing angle must be  $\sim {\cal O}(1)$ to overcome the Yukawa  penalty in the annihilation diagram and yield a thermal annihilation rate. Such large couplings are  
 ruled out by Higgs coupling measurements  and rare meson decay searches \cite{Clarke:2013aya, Krnjaic:2015mbs}. We note that $t$-channel annihilation into a light scalar mediator can still realize secluded annihilation, but this process is less predictive and beyond the scope of this work.

\subsection{Covering  the Thermal Target}
\label{sec:thermaltarget}

 \begin{figure*}[t] 
\hspace{-0.cm}
\includegraphics[width=8.5 cm]{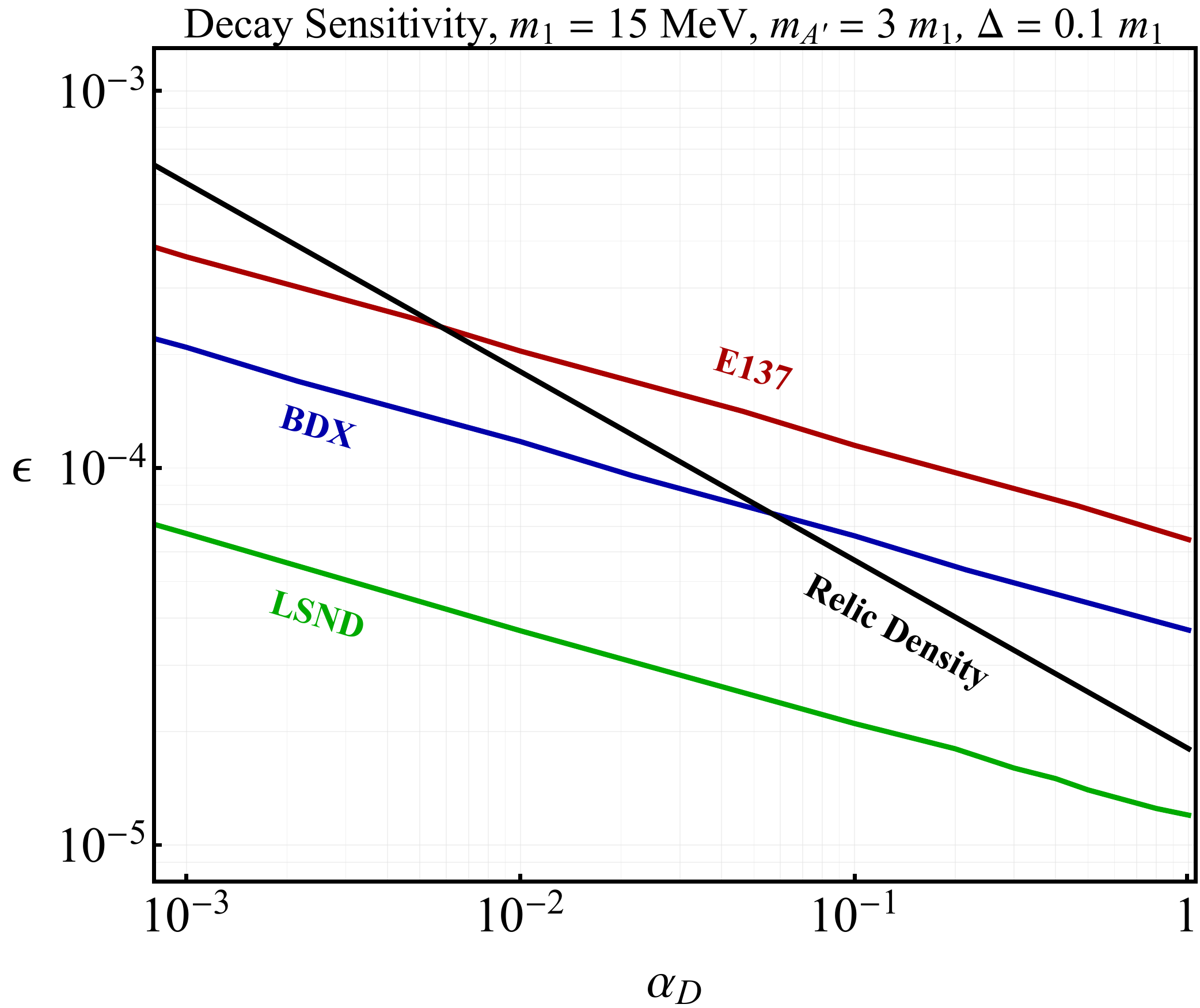}  ~~
\includegraphics[width=8.5 cm]{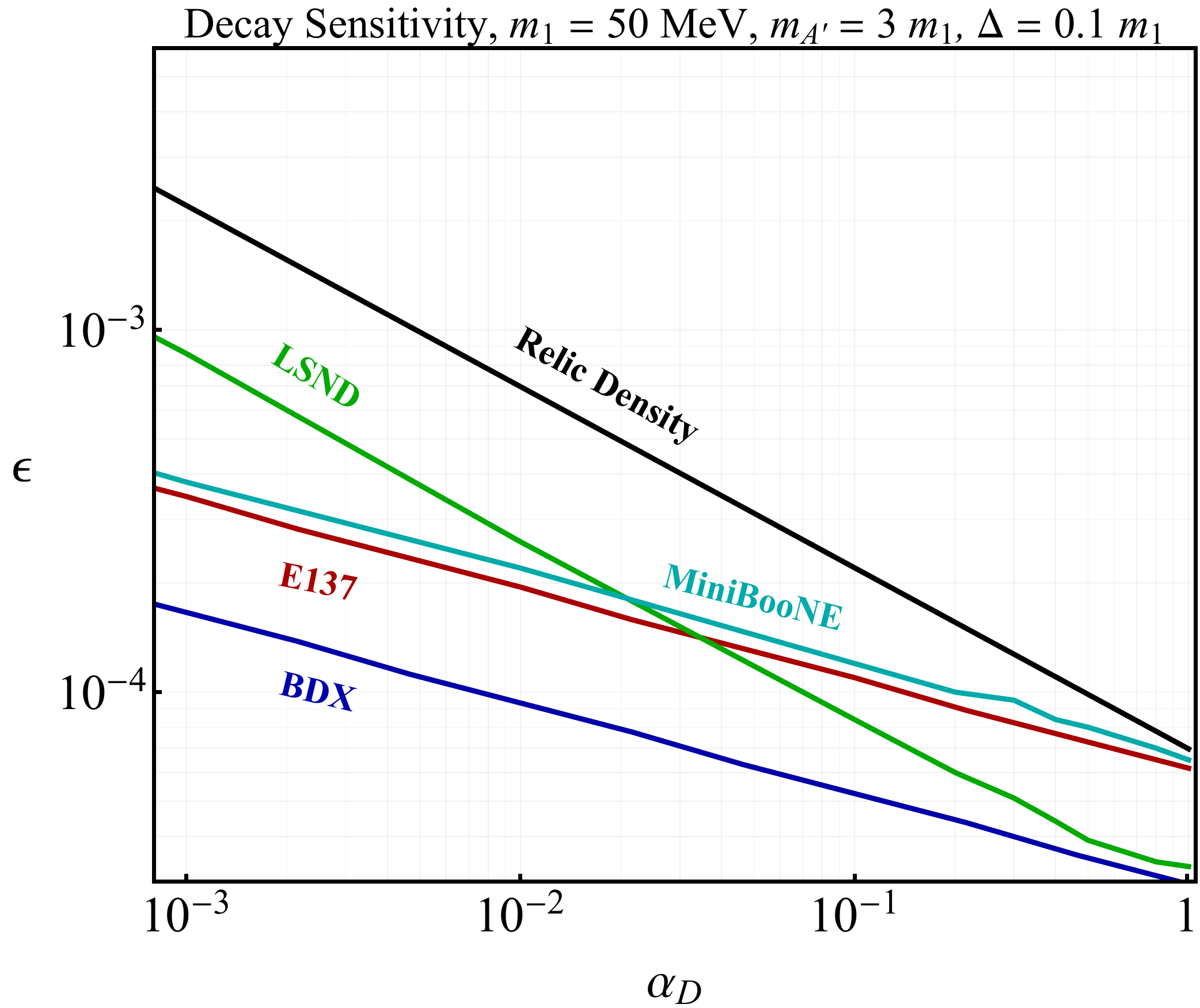} 
  \caption{ Decay reach  in $\epsilon$ as a function of $\alpha_D$, with other model parameters held fixed, for various fixed-target experiments. The value of $\epsilon$ corresponding to the thermal target $y$ for each $\alpha_D$ is shown for comparison in black. The most conservative choice for $\alpha_D$ relative to the thermal target is the largest $\alpha_D$ consistent with perturbativity. Note that the different slope of the LSND curve at 50 MeV compared to the other experiments is a consequence of off-shell $A'$ decay, see Sec.~\ref{sec:protonproduction}. Also note that the MiniBooNE projection from the right panel is missing from the left panel because the mass splitting on the left is 1.5 MeV, which is below the 
 threhsold to pass the experimental cut in Eq.~(\ref{eq:deltamin}). See Sec.~\ref{sec:bounds-projections} for further details on computing the reach.   \label{fig:decaycomparison} }
\vspace{0cm}
\end{figure*}

The goal of this paper is to compute existing bounds on the parameter space for thermal freeze-out via coannihilation and 
present sensitivity projections for future experiments. One major challenge of such an effort is the large dimensionality of 
this parameter space: each viable model point is defined by a unique choice of the inputs $\{m_1, m_{\apr}, \Delta, \epsilon, \alpha_D\}$, constrained
by the requirement  that the coannihilation rate yields the observed DM density. Fortunately, in the nonrelativistic limit, for each choice of $\Delta$, 
the coannihilation rate scales as
\be
\sigma v \propto   \epsilon^2 \alpha_D  \frac{m_1^2}{\, m_{\apr}^4}  = \left( \epsilon^2 \alpha_D \frac{m_1^4}{m_{\apr}^4} \right) \frac{1}{m_1^2} \equiv \frac{y}{m_1^2} ,
\ee
 which is valid for $s \simeq (m_1 + m_2)^2$ and sufficiently away from the $s$-channel
resonance at $m_{\apr} \sim  m_1 + m_2$. Here we have defined the dimensionless parameter 
\be \label{eq:yvariable}
y \equiv  \epsilon^2 \alpha_D    \left( \frac{m_1}{\> m_{\apr} } \right)^4 ,
\ee
which uniquely determines the freeze-out annihilation rate for a given choice of $m_1$ and $\Delta$. 
 
 The virtue of this parameterization is that the observed relic density is achieved only along a one-parameter curve $y(m_1)$ which is insensitive to the relative sizes of $\epsilon, \alpha_D$, or $m_1/m_{\apr}$, reducing the dimensionality of the parameter space. However, the drawback is that some experiments only constrain a subset of model parameters and are not 
sensitive to the same combination of inputs that define the $y$ variable in Eq.~(\ref{eq:yvariable}). For example, lepton 
colliders can be used to constrain $\epsilon$ as a function of $m_{\apr}$ in searches for $e^+e^- \to \gamma {\displaystyle}{\not}{E}$, which
can be interpreted as $e^+e^- \to \gamma \apr$ production followed by prompt $\ap$ decay
to invisible particles \cite{Essig:2013vha,TheBelle:2015mwa,Lees:2017lec}. Since ${\rm Br}(\apr \to {\rm invisible} ) \simeq 1$, the signal event yield 
depends on $\epsilon$, but is insensitive to $\alpha_D$ or the ratio $m_1/m_{\apr}$ for larger $\chi_2$ lifetimes.

Nonetheless, it is still possible to use this information to constrain the $y$ variable for a conservative comparison with the thermal target. 
The strategy is to construct the {\it largest} possible $y_{\rm exp}$ using the experimentally determined $\epsilon_{\rm exp}$, 
\be
         y_{\rm exp} = \epsilon_{\rm exp}^2  \times \left[ \alpha_D  \left(\frac{m_1}{\> m_{\apr}}\right)^4 \right],
\ee
where the quantity in square brackets is chosen to be as large as possible while remaining consistent with both perturbative unitarity 
and the requirement that direct coannihilation remains the dominant annihilation process. Thus, the conservative prescription is to adopt order-one values for $\alpha_D$  and $m_1/m_{\apr}$, while taking $m_{\apr} > m_1 + m_2$ and avoiding the $m_{\apr} \sim m_1 + m_2$ resonance. As an illustration, Fig.~\ref{fig:decaycomparison} shows the reach in $\epsilon$ for $\chi_2$ decay at various experiments, demonstrating that the weakest reach with respect to the thermal target occurs at large $\alpha_D$. For our numerical results in Figs. \ref{fig:money} and \ref{fig:bigmoney}, we chose representative benchmarks $\alpha_D = 0.1, m_1/m_{\apr} = 1/3$ where appropriate (e.g. for
collider bounds). In Fig.~\ref{fig:varyalphaD}  and in the discussion in Sec.~\ref{sec:reach} below, we show how the constrains on the parameter space change for different choices of these benchmark values, and demonstrate that our conclusions are qualitatively unchanged.

Note that for our main results, which are discussed in Sec.~\ref{sec:reach} and presented in Figs.~\ref{fig:money},  \ref{fig:varyalphaD}, and \ref{fig:bigmoney}, we 
compute the $y(m_1)$ curve by numerically solving the full Boltzmann system for this model as described in Appendix \ref{sec:relic}.

\section{Production and Detection Basics}
\label{sec:generic-formalism}
In this section we review the basic $\apr$ production formalism in proton- and electron-beam fixed-target collisions to develop some intuition for our
later numerical results. We also present the formalism for various DM detection signatures that ensue from boosted $\apr \to \chi_1 \chi_2$ decays. 

\subsection{Production at Proton Beam Dumps}
\label{sec:protonproduction}
At proton beam dumps, $\chi_1 \chi_2$ pairs can be produced either from dark bremsstrahlung $p N \to p N \apr$, $\apr \to \chi_1 \chi_2$ or from the neutral meson decays $m^0 \to \gamma A'^{(*)} \to \gamma \chi_1 \chi_2$ where $m^0 = \pi^0, \eta$.\footnote{At sufficiently high energies, direct $A'$ production from quarks and gluons in the target becomes relevant, but this process is negligible for the beam energies we consider in this work.} The meson production rates and spectra depend on various experimental factors, but the strongest dependence is on the proton beam energy, while the detailed material properties of the beam dump can largely be ignored. At few-GeV beam energies, $N_{\pi^0} \approx N_p$ and $N_\eta \approx N_p/30$ \cite{Teis:1996kx} where $N_p$ is the total flux of protons. See Ref.~\cite{deNiverville:2016rqh} for a detailed discussion of production modes and distributions.

If $m_{A'} < m_{m^0}$, the $A'$ can be produced on-shell and the number of $\chi_1\chi_2$ pairs via meson decays is given by
\begin{align}
N_{\chi_1\chi_2} = N_{m^0} \times 2\epsilon^2 \, \left(1 - \frac{m^2_{\apr}}{m^2_{m^0}}\right)^3 \mathrm{BR}(m^0 \to \gamma\gamma) ,
\end{align}
where $N_{m^0}$ is the number of $\pi^0$ or $\eta$ produced.\footnote{If $m_{A'} > m_{m^0}$, the meson decay proceeds through an off-shell $\apr$ and the rate scales as $\epsilon^2 \alpha_D$ \cite{Kahn:2014sra}, in contrast to on-shell production which is independent of $\alpha_D$. At low-energy experiments such as LSND with only a single production channel, off-shell production may be important, but for higher-energy experiments such as MiniBooNE with multiple production channels, on-shell production is dominant.} In either case, there is a kinematic threshold for production through mesons: $m_1+m_2=2m_1 + \Delta < m_{\pi^0}\approx 135\ \mathrm{MeV}$ for pions, and $2m_1 + \Delta <m_{\eta}\approx 548\ \mathrm{MeV}$ for etas. 

The rate of dark bremsstrahlung production is proportional to $\epsilon^2$, but varies non-trivially with the dark photon mass $m_{A'}$. Using the simulation of Ref.~\cite{deNiverville:2016rqh}, we find that the number of $A'$ produced varies from $\sim 10^{-4} \epsilon^2N_p$ as $m_{A'}\to 0$ to $\sim 10^{-9}\epsilon^2N_p$ at $m_{A'}=3\ \mathrm{GeV}$. $A'$ production can be enhanced through vector meson mixing near $m_{A'} = m_\rho, m_\omega$, but since this enhancement is only in a limited mass range and depends on the precise choice of $m_1/m_{\apr}$, we do not attempt to model it precisely. The dark bremsstrahlung production mechanism has no mass threshold and so heavy DM can still be produced up to the beam energy, $2m_1 + \Delta < E_{\rm beam}$. 

\subsection{Production at Electron Beams}
\label{sec:electronproduction}
At electron-beam experiments, mesons are no longer copiously produced so the dominant process is $A'$ production through dark bremsstrahlung followed by on-shell mediator decay $A' \to \chi_1 \chi_2$. The reaction $e N \rightarrow e N \ap$, where 
$N$ is a nucleus of atomic (mass) number $Z(A)$ has been well studied in Refs.~\cite{Kim:1973he,Bjorken:2009mm,Izaguirre:2013uxa}. The relevant features of the reaction can be better illuminated by considering the calculation using the  Weizsacker-Williams (WW) approximation, although we note that all plotted results in this study employ a numerical calculation; see Appendix \ref{app:monte_carlo} for more details of our simulations.

In the WW approximation, the differential production cross section of $\ap$ is then given by
\be
\frac{d\sigma}{dx  \,d \Omega_{\apr}} &\approx& \frac{       4  \alpha^3 \epsilon^2 \, \Phi(q_{0},q_{f})  E^2 x}{ \pi U^2}   \biggl[   \left(1-x+\frac{x^2}{2}\right) \nonumber \\ 
 &&  - \frac{x^2 (1-x) m_{A^\prime}^2  (E^2 \theta^2_{A^\prime}+m_e^2)  }{U^2} \biggr] ,
\ee
where $d\Omega_{\apr} = 2\pi d\cos\theta_{\apr}$ is the  $\apr$ solid angle with respect to the beam axis in the lab frame and we have defined 
\be
& U(x, \theta_{A^\prime}) =  E^2 x \theta^2_{A^\prime} + m_{A^\prime}^2 \frac{1-x}{x} + m_e^2 x,\\
& q_{0} = \frac{2 U}{E(1-x)} \quad, \quad q_{ f} = m_{A'}.
\eea
Here $E$ is the electron beam energy,   $x=E_{A'}/E$ the fraction of the beam energy carried by the $\ap$, and $\alpha\Phi(q_{0},q_{f})/\pi$ is the WW photon flux for small-virtuality photons with momentum $-t$ bounded by $q_{0}^2$ and $ q_{f}^2$  \cite{Kim:1973he}. For $q_{0}$ values much smaller than the inverse nuclear size $\approx 0.4\, \GeV/A^{1/3}$, we have $\Phi(q_{0},q_{f})\sim Z^2$ up to an overall logarithmic factor.  

Ref.~\cite{Bjorken:2009mm} found that for any given $x$, the angular integral is dominated by angles $\theta_{A'}$ such that $E x \theta_{A'}^2 \lesssim  m_{A^\prime}^2 \frac{1-x}{x} + m_e^2 x$. Using this approximation, we can derive  a simpler expression for the differential cross section
\be
\frac{d\sigma}{dx} = 4\alpha^3 \epsilon^2 \overline\Phi(m_{A'},E) \frac{x^2 + 3 x(1-x)}{3m_{A'}^2 (1-x)} ~~ ,\label{dsigmadx}
\ee
where $\overline\Phi(m_{A'},E) \equiv \Phi(q_{0} = m_{A'}^2/(2 E), q_{f} = m_{A'})$.  
After integrating, the total $A'$ production cross-section scales roughly as  
\be\label{crossSectionWW}
\sigma(e N \to eN \apr ) \approx \frac{4\, \alpha^3 \epsilon^2}{3 \, m_{A'}^2} \overline\Phi(m_{A'},E)   \bigl[   \log\delta+{\cal O}(1) \bigr],~~~~~~~
\ee
where $\delta \equiv \min(m_{A'}^2/E^2, m_e^2/m_{A'}^2, m_e^2/E^2)$.

\subsection{Generic Detection Signals (Electron \& Proton Beams)}
\label{sec:generic}
 In our regime of interest, namely $\epsilon^2 \alpha < \alpha_D$, $\ap$ decays promptly primarily into the $\chi_1 \chi_2$ system, imparting an order-one fraction of its energy to the daughter particles, which emerge from the target with large boosts. The boosted DM system gives rise to three classes of observable signatures: detector target scattering,  $\chi_2$ decays, and missing energy/momentum carried away by the DM system. 
 
\subsubsection{Detector Target Scattering} 
\label{sec:scatter-general}
The $\chi_1$ produced in the dump travels unimpeded, enters the detector situated downstream of the target/dump, and scatters off of target particles in the detector. It scatters via the reaction $\chi_1 T \rightarrow \chi_2 T$, where $T$ here could in principle be an electron, a nucleon, or even a nucleus. Similarly, any population of surviving $\chi_2$s produced in the target could downscatter through the reverse reaction $\chi_2 T \rightarrow \chi_1 T$. The cleanest signal occurs when $T$ is an electron, with energy typically above 10s of MeV. The production rate is proportional to $\epsilon^2$ and the scattering rate is proportional to $\epsilon^2 \alpha_D$, so the total yield in this channel will be proportional to $\epsilon^4 \alpha_D$. 

Specializing to the case of electron targets in the  $E_\chi \gg \Delta$ regime, the approximate differential cross section is  \cite{Izaguirre:2013uxa,Batell:2014mga}
\be
\hspace{0.2cm} \frac{d\sigma}{dE_e}  = 4\pi \epsilon^2 \alpha \alpha_D \frac{2 m_e E_\chi^2 - f(E_e) (E_e - m_e)    }{(E_\chi^2 - m_\chi^2)(m_{\apr}^2 + 2m_eE_e - 2m_e^2 )^2},~~~~~~~~~
\ee
where $E_e$ is the electron recoil energy, $E_\chi$ is the incident $\chi_{1,2}$ energy, and we define 
\be
f(E_e)  = 2m_e E_\chi - m_e E_e + m_\chi^2 + 2m_e^2.
\ee
This cross section is valid up to corrections of order $\Delta/E_\chi$ and  $\Delta/m_{\apr}$, both of which are very small compared for the 
benchmarks we consider throughout (see Figs.~\ref{fig:money},  \ref{fig:bigmoney}, and \ref{fig:varyalphaD}). However, our numerical results
 evaluate the exact expression presented following the derivation in \App{sec:DMSMpointlike}.

Neglecting subleading corrections and integrating the recoil energy up to $E_\chi$ gives the approximate result
\be \label{eq:elecscat}
\sigma_{\chi e} \approx \frac{8 \pi \alpha \alpha_D \epsilon^2 m_e E_\chi }{m_{\apr}^4} ~,
\ee
so the corresponding scattering probability inside the detector becomes
\be
P_{\rm scatter} = n_e \sigma_{\chi e} \ell,
\ee
where $\ell$ is the DM path length through the detector and $n_e$ is the electron number density. There
are similar expressions for scattering off detector nucleons and nuclei, but, as we will see, most 
of the relevant bounds and projections below exploit the electron channel. 

 \subsubsection{ Decay of excited state} 
 \label{sec:decay-general}
One of the most powerful channels at beam dump experiments is the direct decay of $\chi_2 \to \chi_1 e^+e^-$, whose partial width satisfies 
\be
\Gamma_{\chi_2}  \approx \frac{4 \epsilon^2 \alpha \alpha_D \Delta^5 }{15 \pi m_{\apr}^4 }  ~~~
\label{eq:approxwidth}
\ee
in the $m_{\apr} \gg m_1 \gg m_e$ limit.\footnote{For $\Delta > 2m_\mu$, $\chi_2$ can also decay to muons, but for the majority of the parameter space we consider in this paper, only the electron channel is allowed.}
The $\chi_2$  is produced in the beam dump and decays in a downstream detector  
(depicted schematically in 
Fig.~\ref{fig:beam-dump-cartoon}), so the signal yield scales as 
\be 
N_{\rm signal} \propto  \epsilon^2 P_{\rm survive} P_{\rm decay},
\ee
where the product  
\be \label{eq:probs}
 P_{\rm survive} \,   P_{\rm decay} = e^{-\Gamma_{\chi_2} d/\beta\gamma} \left(1 - e^{-\Gamma_{\chi_2} \ell/\beta\gamma} \right)
 \ee
is the probability for $\chi_2$ to survive a distance $d$ out to do the detector and decay inside after traversing a path length $\ell$ with
boost factor $\gamma$ and velocity $\beta = v/c$. As with scattering, the total decay yield scales as $\epsilon^4 \alpha_D$, with $\epsilon^2$ coming from production and $\epsilon^2 \alpha_D$ coming from expanding the exponentials in Eq.~(\ref{eq:probs}) assuming the long lifetime limit (see Fig.~\ref{fig:decaylength}) and using Eq.~(\ref{eq:approxwidth}).

To estimate the relative reach of decay and scattering searches at  beam dump
experiments, it is useful to define
 \be
R \equiv \frac{ P_{\rm decay}}{P_{\rm scatter}} \simeq \frac{\Delta^5}{60 \pi^2 \gamma  E_\chi m_e n_{e}} ,
 \ee
where $n_e$ is the number density of detector electrons. Here we have used 
the approximate $\chi$-electron scattering cross section from Eq.~(\ref{eq:elecscat}); 
note that this expression is independent of detector geometry or 
DM production rate. 

For most materials, $n_e  \sim 10^{24} \ \cm^{-3}$, so we have 
\be 
R \simeq 3   \left(  \frac{m_1}{ 20 \, \MeV} \right)   \left(  \frac{\Delta}{0.1 m_1} \right)^5
\left(  \frac{10 \, \GeV}{ E_\chi} \right)^2,
\ee
where $E_\chi$ represents the typical energy of a $\chi$ within the detector acceptance. Note that the 
$R\sim \Delta^5$ scaling implies that decays dominate the reach for most of the splittings shown in 
Fig.~\ref{fig:money}, and that for experiments with 
lower DM energies (e.g. at LSND where $E_\chi \sim$ few 100 MeV) the decay reach
dominates the scattering reach by several orders of magnitude due to the lower beam energy and smaller boost factors. 

 \subsubsection{ Upscatter followed by decay}
 \label{sec:upscatter-and-decay}
 A third possible signal combines the phenomenology of both scattering and decay. $\chi_1$ can upscatter at the detector, producing a $\chi_2$ along with a recoiling target $T$. If the $\chi_2$ produced in the upscatter decays inside the detector via the reaction $\chi_2 \rightarrow \chi_1 e^+ e^-$, and if the electron and positron final states are energetic and separated enough, the final state leads to a signature that is not easily mimicked by environmental or beam-related backgrounds \cite{Izaguirre:2014dua,Battaglieri:2014qoa,Battaglieri:2016ggd}. However, we find that the yields in this channel are subdominant to the other two and we will not consider it further.\footnote{If the recoiling target $T$ is not visible, the signal in this channel is identical to the decay signal, but with the added $\epsilon^2 \alpha_D$ penalty of scattering.} The consequences of this signal for the case of boosted astrophysical DM, where the different kinematics and the absence of an $\epsilon^2$ production penalty make this channel an attractive background-free option, were investigated in detail in Ref.~\cite{Kim:2016zjx}.

 \subsubsection{Missing Energy/Momentum}
 \label{sec:mme-general}
 A unique feature of $\apr$ production in electron-beam fixed-target collisions is that the $\apr$ is typically radiated with
 an order-one fraction of the incident beam energy (see Sec.~\ref{sec:electronproduction} and more detailed discussions in
 \cite{Kim:1973he,Bjorken:2009mm,Izaguirre:2014bca,Banerjee:2016tad}). This enables a novel detection strategy where the target is now embedded in the detector, and which is
based on comparing the energy and momentum of the beam electron before and after it undergoes dark bremsstrahlung. Fig.~\ref{fig:mme-cartoon} illustrates a set-up for this detection strategy. 

\subsection{Kinematic Thresholds}

There are several kinematic thresholds which influence the detectability of the various signals. Clearly, for $\Delta< 2m_e\approx 1\ \mathrm{MeV}$, the excited state $\chi_2$ can no longer decay via $\chi_2\to\chi_1\epm$. The only possible decay mode remaining is to $\chi_1$ plus neutrinos or $3\gamma$, both of which are sufficiently suppressed that $\chi_2$ is cosmologically long-lived \cite{Batell:2009vb}. In this case the only signals are from the recoiling target in $\chi_1$ upscattering or $\chi_2$ downscattering, along with missing energy/momentum. Otherwise, for detectors with finite energy thresholds and angular resolution, we might require the decay signal $\chi_2\to\chi_1\epm$ to have an energetic, well-separated $\epm$ pair satisfying $E_e>E_{\rm min}$ and $\theta_{\epm}>\theta_{\rm min}$. The requirement on $\Delta$ to allow a visible $\chi_2$ decay is actually more stringent:
\be
\Delta^2 \geq 2m_e^2\left(1+\cos\theta_{\rm min}\right) + 2E_{\rm min}^2\left(1-\cos\theta_{\rm min}\right).
\label{eq:deltamin}
\ee
For example, requiring $E_{\rm min} = 50 \ \MeV$ and $\theta_{\rm min} = 2^\circ$ as at MiniBooNE, we find that the minimum splitting we can probe is $\Delta \gtrsim 2\ \mathrm{MeV}$.
 
\section{Existing Bounds and Projections}
\label{sec:bounds-projections}

In this section we discuss the key features of the various representative experiments we consider and list our kinematic cuts used to compute reach curves. The bounds and projections we derive are presented in Figs. \ref{fig:money}, \ref{fig:varyalphaD}, and \ref{fig:bigmoney}.

\subsection{Signals at Proton Beam Dump Experiments }
\subsubsection{LSND}
The Liquid Scintillator Neutrino Detector (LSND) \cite{Athanassopoulos:1996ds} was a neutrino experiment at Los Alamos which ran from 1993-1998. The extremely high-luminosity proton beam produced the largest available fixed-target sample of neutral pions, $N_{\pizero}\sim 10^{22}$. The LSND proton beam had kinetic energy 800 MeV, small enough that $\eta$ production and dark bremsstrahlung are negligible. Therefore, the DM is produced dominantly from $\pi^0$ decays. We modeled the $\pizero$ production at LSND using the GEANT simulation from Ref.~\cite{Kahn:2014sra} and then decayed the pions to DM using Monte Carlo. Due primarily to the large luminosity, we expect DM signal event yields at LSND to dominate in the regions of parameters space where pion decay is kinematically allowed.

The LSND detector was an approximately cylindrical tank of scintillating mineral oil with length $\sim 10\mathrm{m}$ and radius $\sim 3\mathrm{m}$, located $\sim 30\mathrm{m}$ from the beam stop. The close proximity to the beam stop and its off-axis orientation gives a large geometric acceptance of tracks originating at the beam stop. The detector used photomultiplier tubes to detect Cerenkov light emitted by charged particles in the scintillator. For the purposes of our simulations, LSND is only capable of identifying lepton tracks but is blind to nucleons, and electrons are indistinguishable from positrons. We take the electron detection efficiency at LSND to be $19\%$ \cite{Auerbach:2001wg}.

We numerically computed the event yields at LSND using DM events that were produced as outlined in Sec.~\ref{sec:protonproduction}, for each of the available signal channels in Sec.~\ref{sec:generic}. Using the techniques outlined in App.~\ref{app:DMdetection}, we simulated the $\chi_1$ and $\chi_2$ scattering off electrons, nucleons, and nuclei as well as direct $\chi_2$ decays in the detector. This simulation accounted for the geometric acceptance of the detector as well as kinematic cuts and detection efficiencies for the observable final state particles.

Previous dark matter limits at LSND have been derived using the 55-event background-subtracted limit from the LSND neutrino-electron elastic scattering analysis \cite{Auerbach:2001wg,deNiverville:2011it}. To derive the LSND decay constraints for inelastic DM, we interpreted the 55-event limit using $\chi_2$ decay events where the $e^+/e^-$ opening angle was too small to distinguish the two leptons, and thus could fake an elastic event passing the cuts described in Ref.~\cite{Auerbach:2001wg}. The angular resolution of LSND is $12^{\circ}$ for electrons above 20 MeV \cite{Athanassopoulos:1996ds}, so we require that the angular separation of the $\epm$ pair is $\theta_{\epm}<12^{\circ}$ and that the total energy of the pair lies in the range $18<E_{e^+}+E_{e^-}<50\ \mathrm{MeV}$ with either the electron or positron satisfying $\cos\theta<0.9$. This is conservative, as we have ignored a similar number of events where the electron and positron are well-separated; with access to the full LSND data set, one could search for events where two leptons were seen simultaneously inside the detector and potentially improve the reach. Indeed, we chose such a conservative prescription because, as we will see, the limits derived from LSND are already dominant for $\Delta > 2m_e$, $2m_1 + \Delta < m_{\pi^0}$. 

For the LSND scattering constraints, we again use the 55-event limit applied to any event with \emph{at least} one lepton passing the elastic cuts in the final state. This includes $\chi_2$ down-scattering and $\chi_1$ upscattering with and without $\chi_2$ decay, though the dominant process for all but the largest DM masses is $\chi_2$ downscattering, for which the signal is identical to neutrino-electron elastic scattering. We require a recoiling electron track to be forward-oriented with $\cos\theta>0.9$ and have energy between $18<E_e<50\ \mathrm{MeV}$. Note that because LSND does not have particle ID, a potential background is neutral-current $\pi^0$ production where the photons from the $\pi^0$ decay fake electrons. 

\subsubsection{MiniBooNE}

MiniBooNE \cite{AguilarArevalo:2008qa} is a proton beam neutrino experiment currently operating at Fermilab with beam energy 8 GeV. It has been previously noted that MiniBooNE is sensitive to light DM \cite{deNiverville:2011it,deNiverville:2012ij}, and indeed, the first limits from a dedicated DM search have recently been published \cite{Aguilar-Arevalo:2017mqx}. At MiniBooNE, the beam energy is large enough that $\eta$ production and dark bremsstrahlung contribute to DM production, however, contributions from $\pizero$ decays dominate in regions of parameter space where the $\pizero\to\gamma\chi_1\chi_2$ is allowed. Due to the smaller luminosity, the number of pions produced at MiniBooNE is only $N_{\pizero}\sim 10^{20}$, with the number of etas being further suppressed by a factor of $\sim 30$. We used \verb|BdNMC| \cite{deNiverville:2016rqh} to generate a distribution of $\pizero$ and $\eta$ produced at MiniBoone and then decayed the mesons to DM using Monte Carlo. To simulate dark bremsstrahlung production at MiniBooNE, we used \verb|BdNMC| to generate a distribution of on-shell $A'$ and then decayed the dark photons to DM using Monte Carlo. The rate of dark bremsstrahlung production decreases as $m_{A'}$ is increased, but for $m_{A'} \gtrsim 200\ \mathrm{MeV}$, kinematic thresholds and $\alpha_D$ suppression from off-shell meson decay are significant enough that the dark bremsstrahlung production of DM dominates. A more detailed discussion of the Monte Carlo methods and matrix elements used in these simulations can be found in Appendices \ref{app:matrix_elements} and \ref{app:monte_carlo}.

The MiniBooNE detector also uses scintillating mineral oil, and is approximately spherical with radius $\sim 5\mathrm{m}$, located $\sim 500\mathrm{m}$ from the beam stop. The smaller detector and larger distance from the beam stop means that the geometric acceptance is smaller than LSND, but because the beam energy of MiniBooNE is ten times that of LSND, the DM produced at MiniBooNE has boost factors that are roughly ten times greater, which to some extent compensates for the geometric acceptance as the DM is more collimated. That said, the combination of the smaller proton luminosity and the larger number of background events for a DM search \cite{Aguilar-Arevalo:2017mqx} results in the MiniBooNE scattering reach being suppressed by several orders of magnitude compared to LSND. We do not include the MiniBooNE scattering curves in our reach plots as they do not cover new parameter space.\footnote{SBND \cite{Antonello:2015lea} uses the same beamline as MiniBooNE, but with a smaller detector placed closer to the beam stop. We expect the decay reach to be similar to MiniBooNE for all but the largest mass splittings, but the scattering reach should be enhanced due to the higher detector efficiency \cite{VanDeWaterPrivate}.} MiniBooNE is capable of seeing both nucleon and lepton tracks, but electrons are indistinguishable from positrons and neutrons are indistinguishable from protons.\footnote{However, MiniBooNE can distinguish photons from leptons \cite{Patterson:2009ki}.}  We take the efficiency for lepton and nucleon detection to be $35\%$ \cite{Patterson:2009ki}.

At MiniBooNE, we impose energy and angular cuts similar to those in Ref~\cite{Aguilar-Arevalo:2017mqx}. We require any lepton track to be forward-oriented with $\cos\theta>0.99$ and have energy in the range $50 < E_e < 600\ \mathrm{MeV}$. We also require the electron and positron to have an angular separation of at least $2^{\circ}$ \cite{Patterson:2009ki} and an energy of at least $50\ \mathrm{MeV}$ to be visible. We determined the decay reach assuming 95\% one-sided Poisson c.l., corresponding to 3 events.
\subsection{Signals at Electron Beam Dump Experiments}
\subsubsection{E137}
The E137 experiment at SLAC \cite{Bjorken:1988as} was designed to search for axion-like particles and 
with a 20 GeV beam delivering $\sim$ 30 C ($\sim 2 \times 10^{20}$ electrons) of current onto an aluminum target positioned approximately 400 m upstream from an aluminum and plastic scintillator detector. Ref. \cite{Batell:2014mga} demonstrated that the existing data from this sample is sensitive to sub-GeV DM if the DM 
can be produced in the beam dump and scatter {\rm elastically} off detector electrons. 

The null result of the E137 search can be used to place bounds on our inelastically coupled scenario. The DM yield
is computed by evaluating the $\chi$ flux produced in the beam dump via dark-brehmstrahlung (described in Sec. \ref{sec:electronproduction}) and
considering both DM scattering off electron targets in the detector (described in Sec.~\ref{sec:scatter-general}), as well as $\chi_2$ decays inside the detector volume (Sec.~\ref{sec:decay-general}). To comply with the 
search criteria from Ref.~\cite{Bjorken:1988as}, we demand either signal process deposit at least 1 GeV of energy inside the detector's geometric acceptance, and
we place a 3-event bound to account for the null result of the E137 experiment.

\bigskip

\subsubsection{BDX}
 The proposed BDX experiment at Jefferson Laboratory \cite{Battaglieri:2014qoa,Battaglieri:2016ggd,Izaguirre:2013uxa,Izaguirre:2014dua}
is a dedicated effort to search for light DM produced and detected in analogy with the procedure at E137.\footnote{A similar idea 
involving lower-energy electron beams installed near large underground neutrino detectors was proposed in Ref.~\cite{Izaguirre:2015pva}.} The setup 
involves placing a meter-scale CsI scintillator detector behind the beam dump at the upgraded 11 GeV CEBAF beam.
In comparison with E137, BDX has greater luminosity,
($\sim 10^{22}$ electrons on target), a shorter
baseline ($\sim 20$ m distance from dump to detector), and a larger detector volume ($50 \times 50 \times 200 \, \cm^3$). 

In the BDX setup, the boosted $\chi_1\chi_2$ system emerges from the beam dump
and can either scatter (see Sec.~\ref{sec:scatter-general}) via $\chi_i e \to \chi_j e$ or the excited state can decay via $\chi_2 \to \chi_1 e^+e^-$ Sec.~\ref{sec:decay-general}) 
as it passes through the $\sim 2$ meter long detector. We compute decay and scattering projections using 3 and 10 event yield contours, respectively, for EM energy depositions above 300 MeV.
 
\subsection{Signals at Electron Missing Momentum/Energy Experiments}
\label{sec:electron-missing-momentum}

\subsubsection{NA64}
\label{sec:NA64}
The NA64 experiment at the CERN SPS \cite{Banerjee:2016tad}, depicted schematically in Fig.~\ref{fig:mme-cartoon} (right panel), is sensitive to DM production via 
dark bremsstrahlung as described in Sec.~\ref{sec:electronproduction}. In this setup,  $2 \times 10^9$ electrons with 100 GeV energies are 
fired into an active ECAL target which measures the  energy/momentum and triggers on events with large missing energy. In principle,
this technique is sensitive to our scenario of interest because the excited state can decay outside the detector, thereby giving rise to 
missing energy in the ECAL measurement. While we include this 
discussion for completeness, we find that the existing data sample does not currently constrain new parameter space, so we do not discuss it further. 

\subsubsection{LDMX}
\label{sec:LDMX}
The proposed LDMX experiment at SLAC \cite{Izaguirre:2014bca,Alexander:2016aln} aims to produce DM using the 4 or 8 GeV LCLS-2 electron beam passing through 
 a thin target upstream of a dedicated tracker and ECAL/HCAL system designed to veto SM particles produced in these collisions -- the setup is depicted schematically in Fig.~\ref{fig:mme-cartoon} (left panel). For our inelastically  coupled DM scenario, an $\apr$ is produced 
 via dark bremsstrahlung (described in Sec.~\ref{sec:electronproduction}) and decays via $\apr \to \chi_1\chi_2$ followed by a displaced 
 $\chi_2 \to \chi_1e^+e^-$ decay. If this decay occurs behind the ECAL/HCAL system, it can mimic the missing energy signature which LDMX is optimized to observe. 
 
 A representative realization of this setup involves $\sim 10^{16}$ electrons with 4 GeV energies passing through a tungsten target of thickness $\sim 0.1 X_0$ and emerging from
 the target with less than 1 GeV of energy. The signal yield scales as 
 \be
 N_{\rm signal} \approx  N_{e^-} n_{W} \sigma(eN\to eN\apr) X_0 e^{-\Gamma_{\chi_2} \ell / \beta\gamma}~,
 \ee
 where $N_{e} = 10^{16}$ is the number of  electrons on target,  $X_0 \sim 9$ cm is the tungsten radiation length,  $n_W$ is the 
 tungsten number density,
$ \sigma(eN\to eN\apr) $ is the $\apr$ production cross section from Eq.~(\ref{crossSectionWW}) and  $\beta/\gamma$ are the 
$\chi_2$ velocity and boost factors, respectively. Here the exponential factor is the $\chi_2$ survival probability through 
a path length of $\ell$ through the LDMX ECAL/HCAL system. The LDMX missing momentum projections present the 3 event yield contours for various parameter benchmarks. 

We note in passing that it may also be possible for both NA64 and LDMX to be sensitive to a promptly decaying $\chi_2 \to \chi_1 e^+e^-$
inside their ECAL systems, but the backgrounds for this process are not known and understanding this signal is beyond the scope of the present work.

\section{Plots  \& Main Results }
\label{sec:reach}

\begin{figure*}[t] 
\hspace{-0.5cm}
\includegraphics[width=8.5cm]{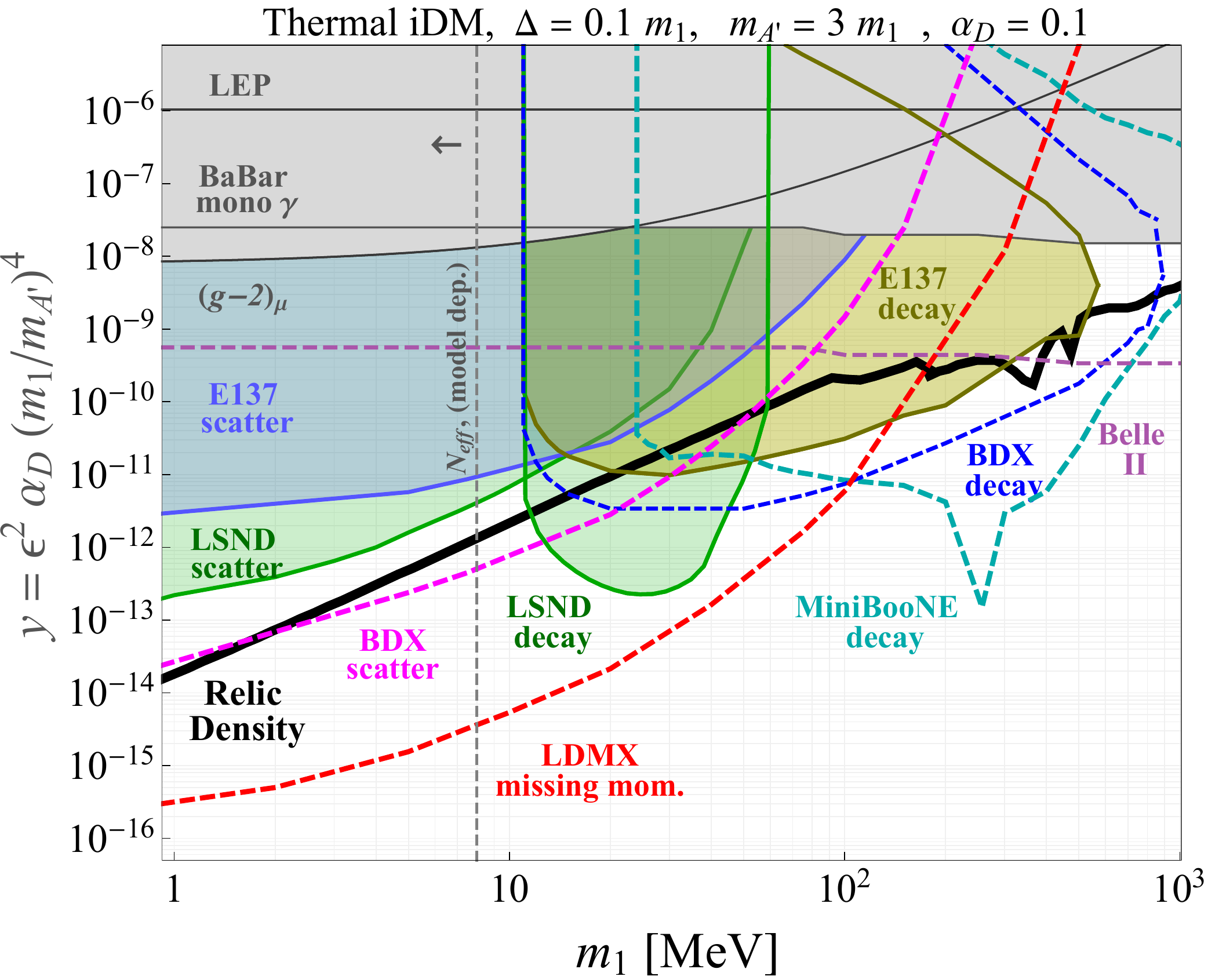}  ~~
\includegraphics[width=8.5cm]{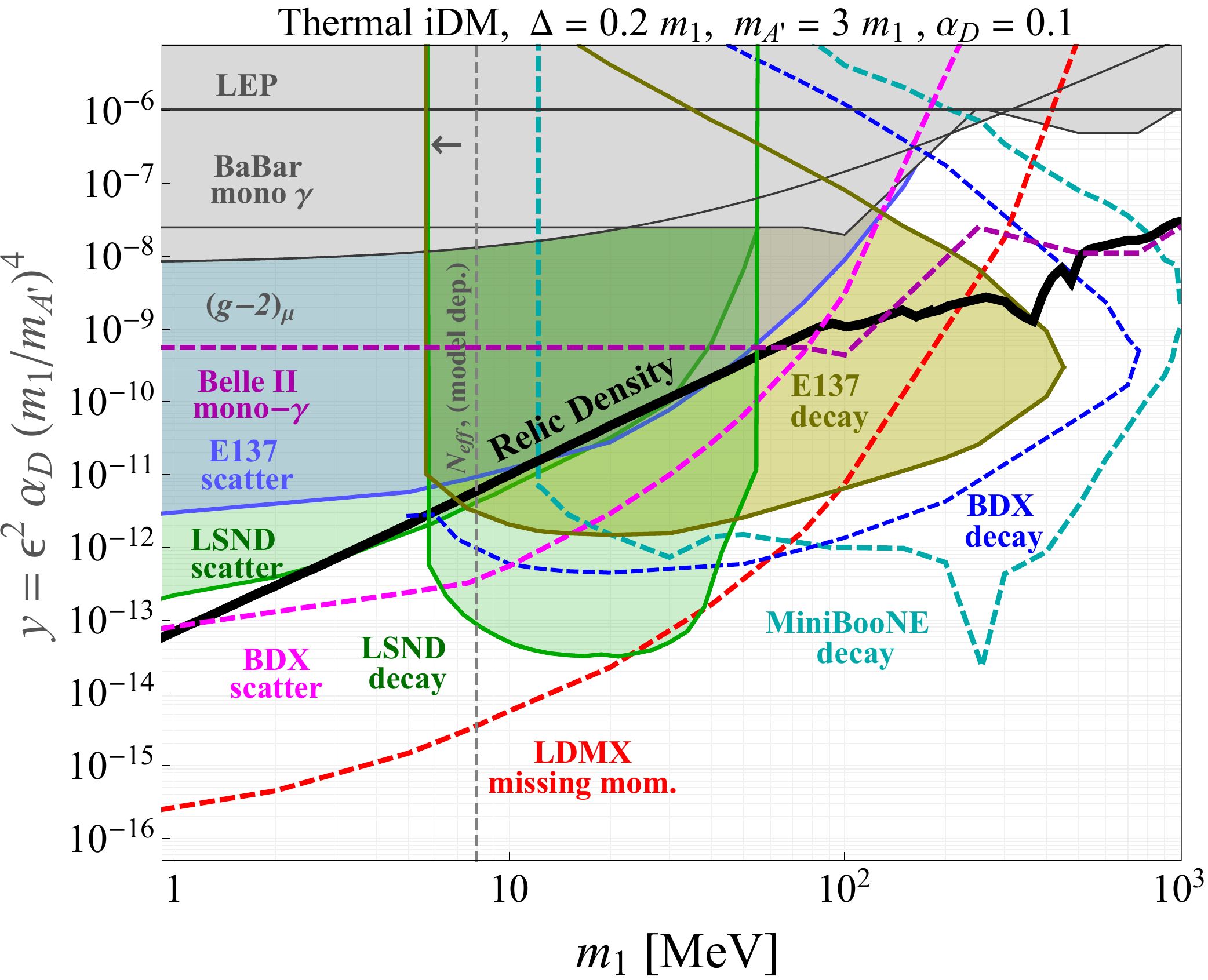}  ~~
\includegraphics[width=8.5cm]{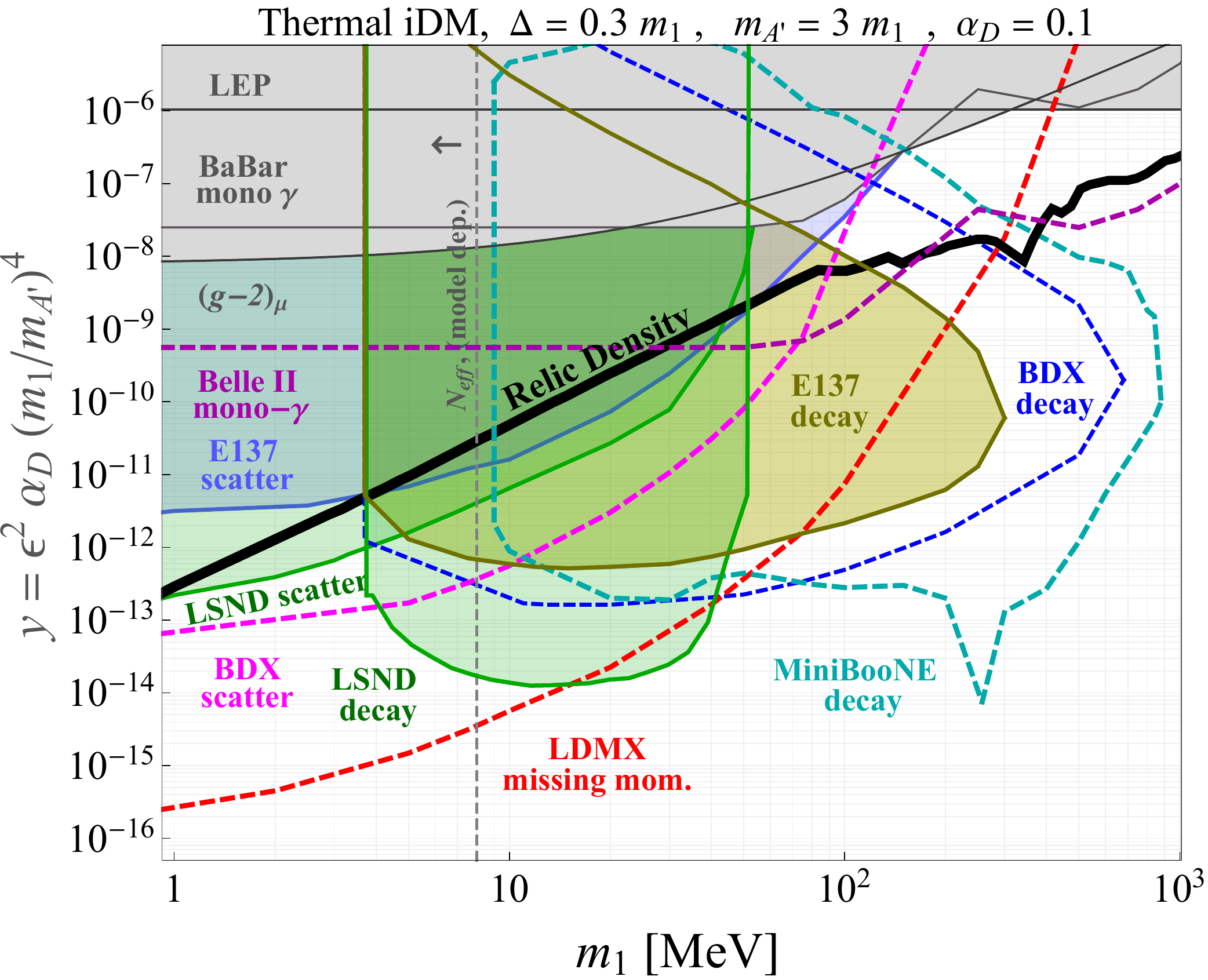}  ~~
\includegraphics[width=8.5cm]{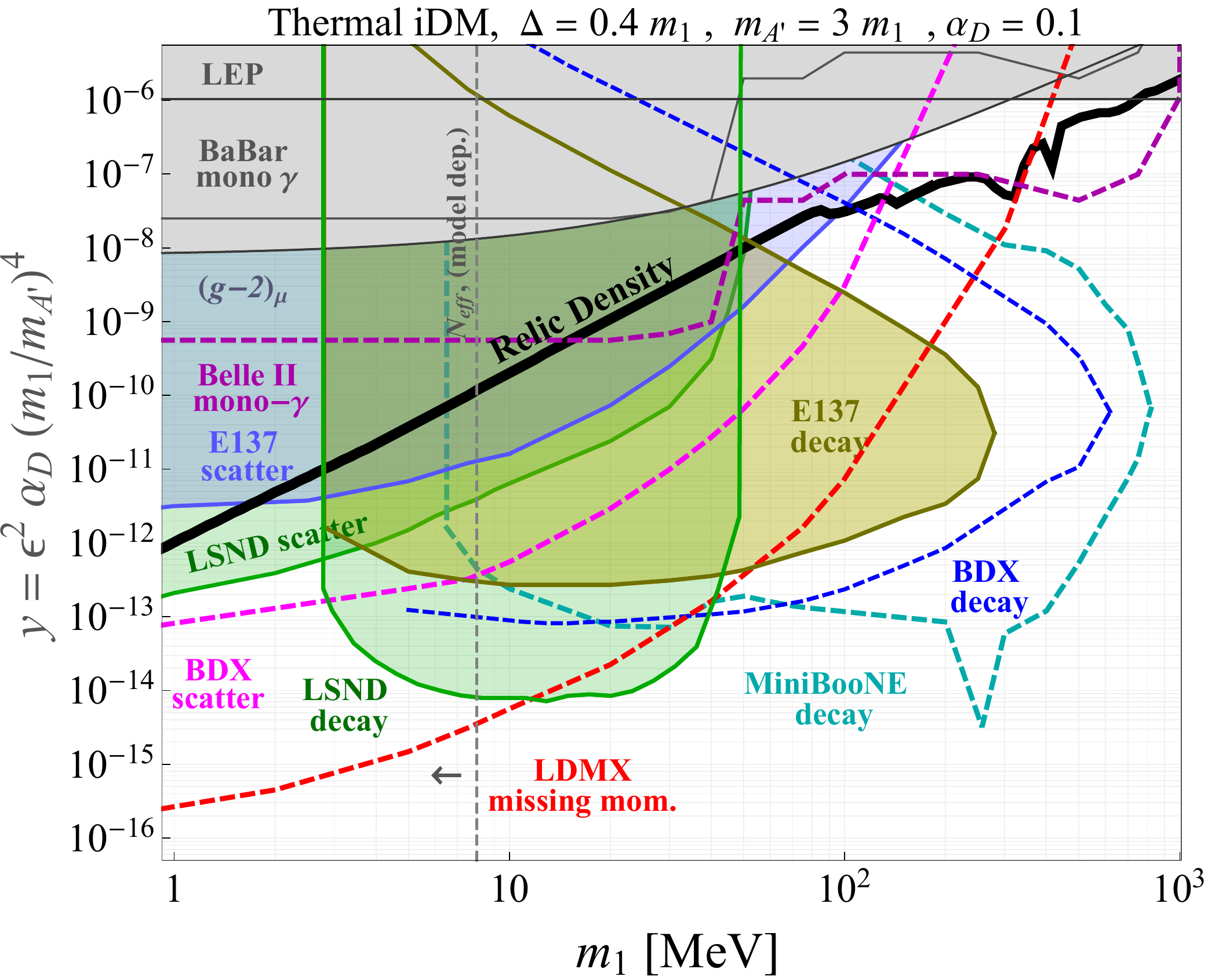}  ~~
  \caption{Parameter space compatible with thermal inelastic DM for different choices of $\Delta$ with constraints and future projections presented. The black relic density curve is computed using the procedure described in Appendix \ref{sec:relic}.
  For each choice of $\Delta$, the relic density curve is insensitive to the relative values of $\epsilon, \alpha_D$, or $m_1/m_{\apr}$, however, some
  constraints depend sensitively on these choices. Typical examples of this sensitivity are LEP and $(g-2)_\mu$ for which
  the curves shown here are based only on their limits on $\epsilon$; the observables in question do not depend on $\alpha_D$ or the DM/mediator mass ratio. Thus, where appropriate, we have adopted the conservative prescription $\alpha_D = 0.1$ and $m_{\apr}/m_1 = 3$ to place these constraints
  on this plot, thereby revealing the remaining viable parameter space; see text for a discussion.  The colored curves in these plots represent new results computed in work: solid lines are existing constraints, and dashed lines are projections. The Belle II \cite{TheBelle:2015mwa} projections are estimated by rescaling the BaBar luminosity for the process $e^+e^- \to \gamma \apr \to \gamma \chi_1 \chi_2$  in which the $\chi_2$ decays outside the detector. The gray shaded regions represent kinetic mixing constraints $(g-2)_\mu$  \cite{Pospelov:2008zw}; LEP \cite{Hook:2010tw}; and BaBar \cite{Izaguirre:2015zva}. Finally, the vertical dashed line labeled $N_{\rm eff.}$ is a 
  model-dependent bound from DM freeze-out reheating photons preferentially over neutrinos during BBN \cite{Nollett:2013pwa}, excluding parameter space to the left of the line; if there are
  other sources of dark radiation, this bound can be alleviated.}
   \label{fig:money}
\vspace{0cm}
\end{figure*}

\begin{figure*}[t] 
\hspace{-0.6cm}
\includegraphics[width=8.7cm]{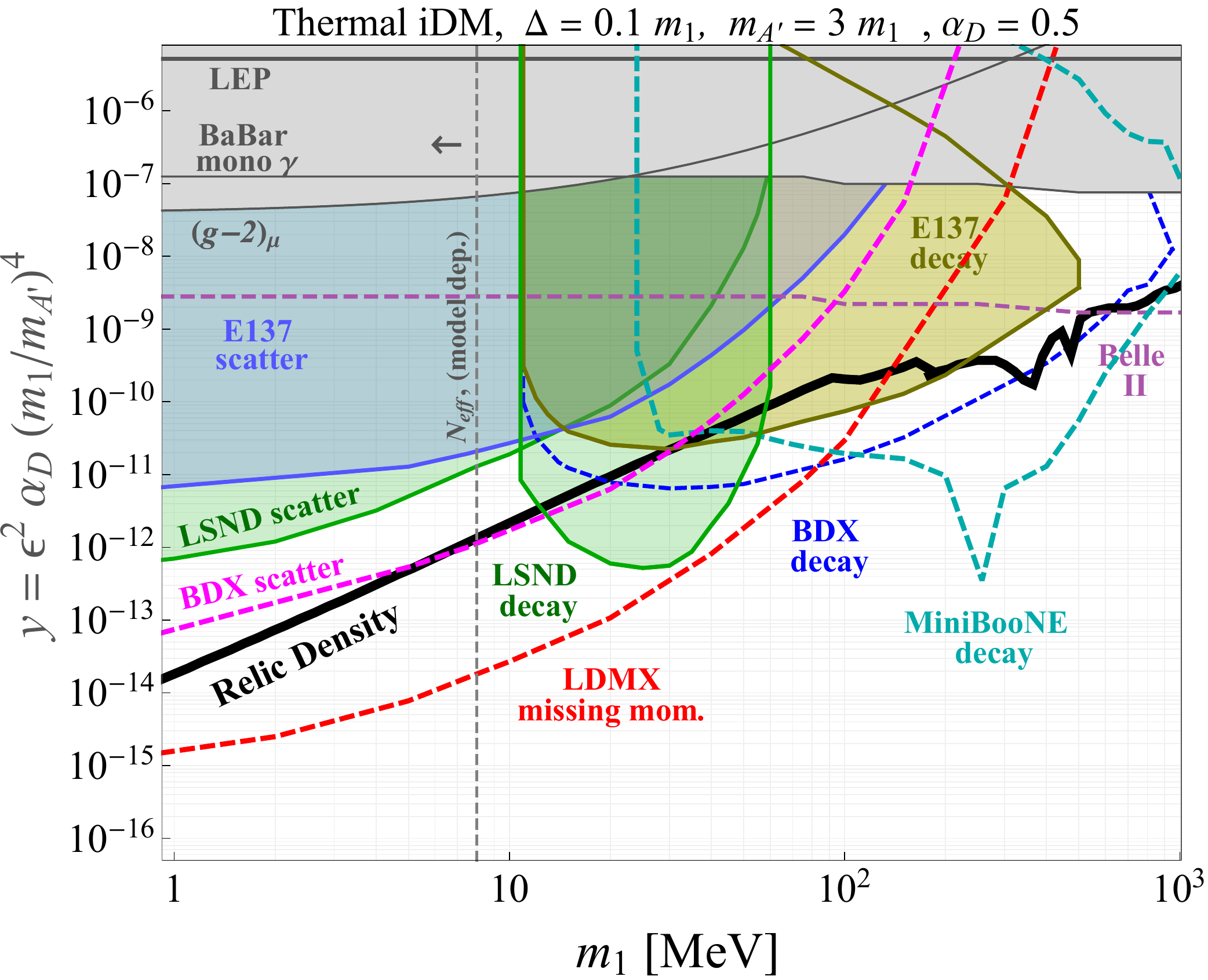} ~~
\includegraphics[width=8.7cm]{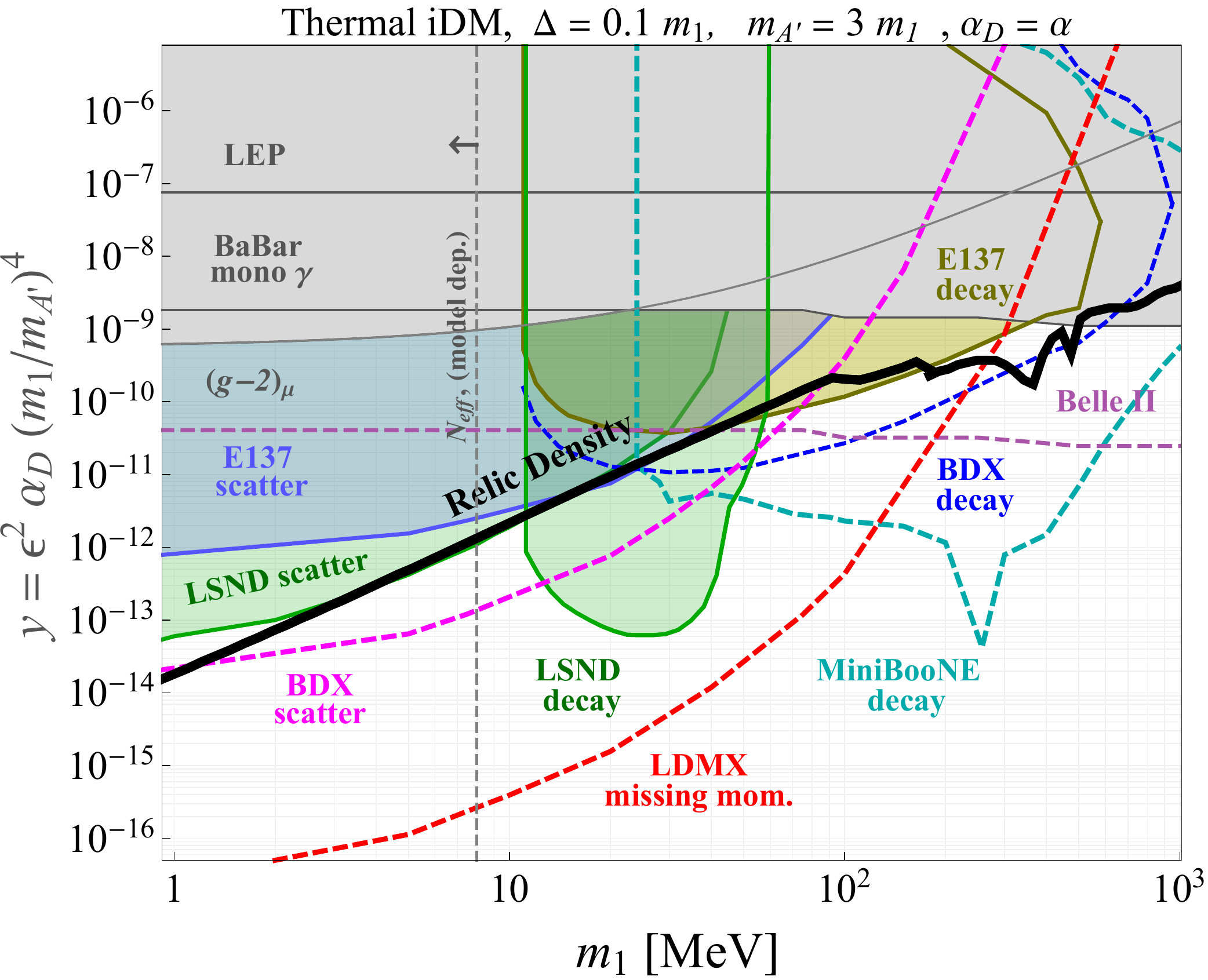} ~~ \\
\hspace{-0.6cm} \includegraphics[width=8.7cm]{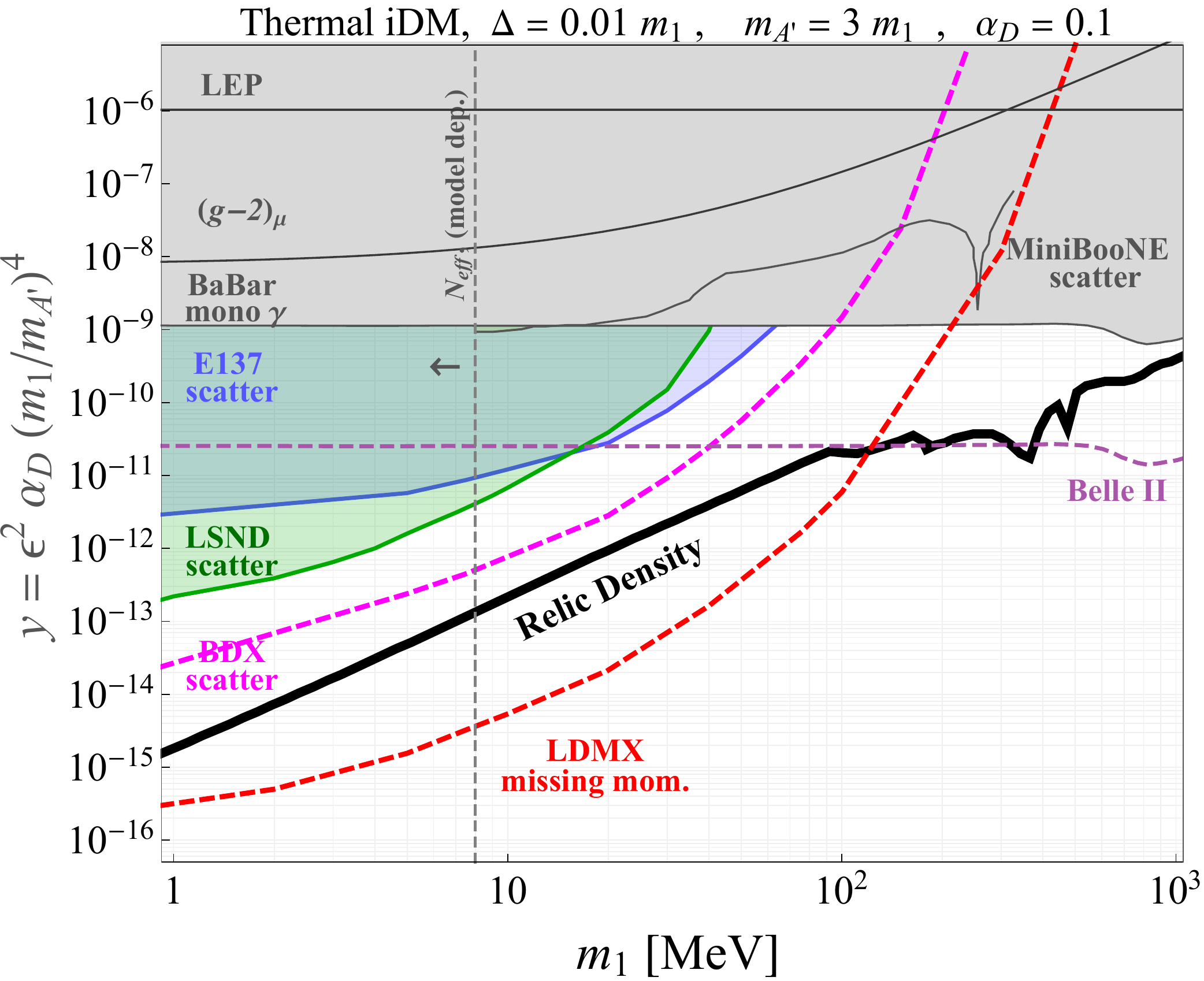}  ~~
\includegraphics[width=8.7cm]{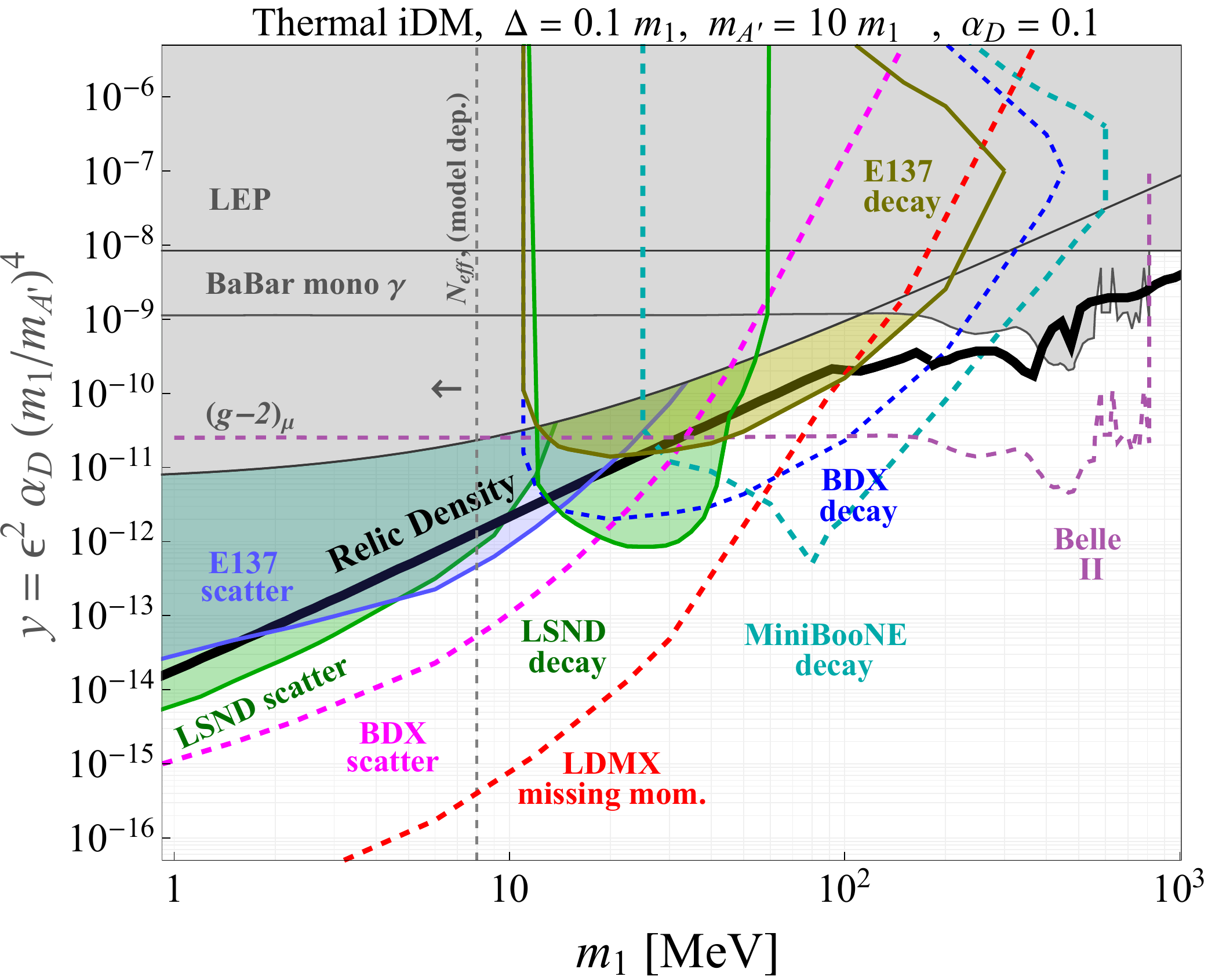}  ~~
  \caption{{\bf Top row:} Same as the top-left panel of Fig.~\ref{fig:money}, but with different choices for $\alpha_D$. For larger couplings near the perturbativity limit \cite{Davoudiasl:2015hxa} (left with $\alpha_D = 0.5$) the viable parameter space increases slightly relative to Fig.~\ref{fig:money}.  
  For smaller couplings (right with $\alpha_D = \alpha$) the thermal target is nearly closed. {\bf Bottom row: } Same as the-top left panel in Fig.~\ref{fig:money}, 
  but with different choices for the inelastic splitting $\Delta = 0.01 m_1$ (left) and the mediator mass $m_{\apr} = 10 m_1$ (right). Note that
  for very small mass splittings, the decay searches become ineffective and the best limits arise from scattering and collider searches, whose observables
  do not rely on a prompt $\chi_2$ decay.}
  \label{fig:varyalphaD}
  \vspace{0cm}
  \end{figure*}
  
  \begin{figure*}[t] 
\includegraphics[width=17cm]{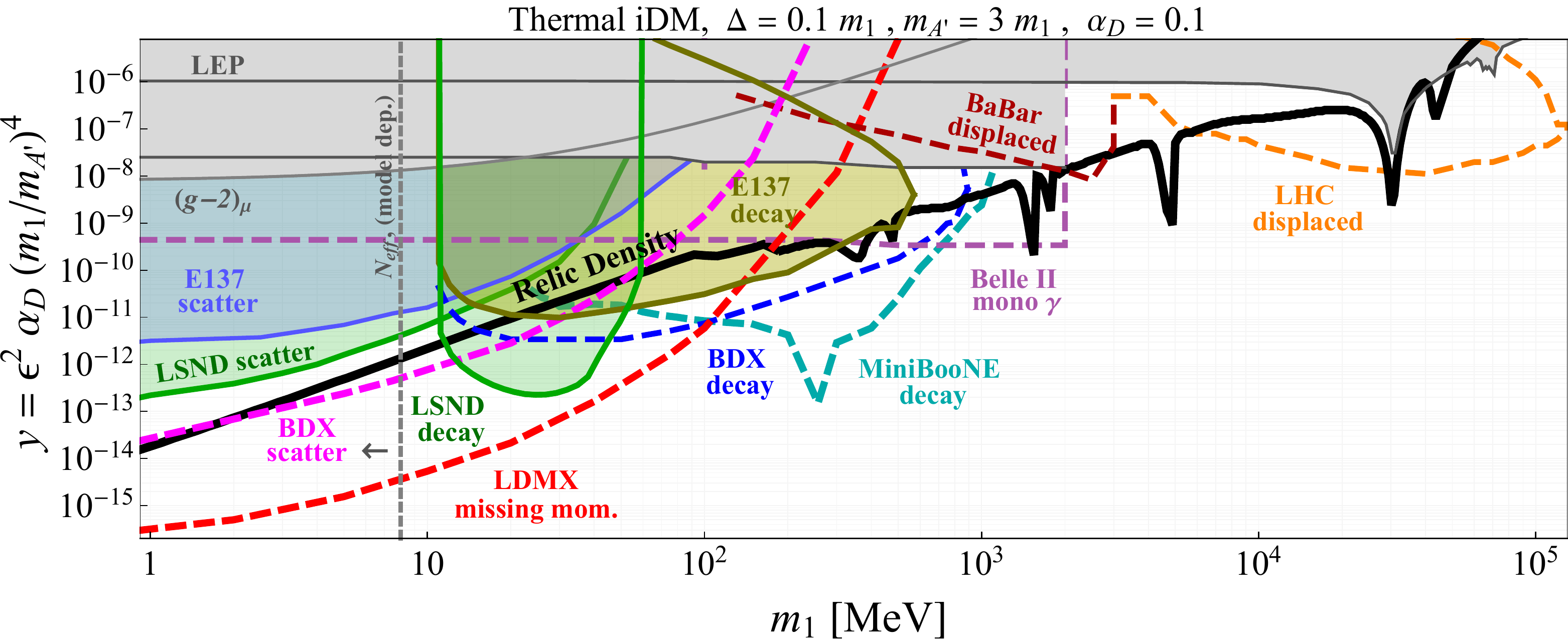} 
\includegraphics[width=17cm]{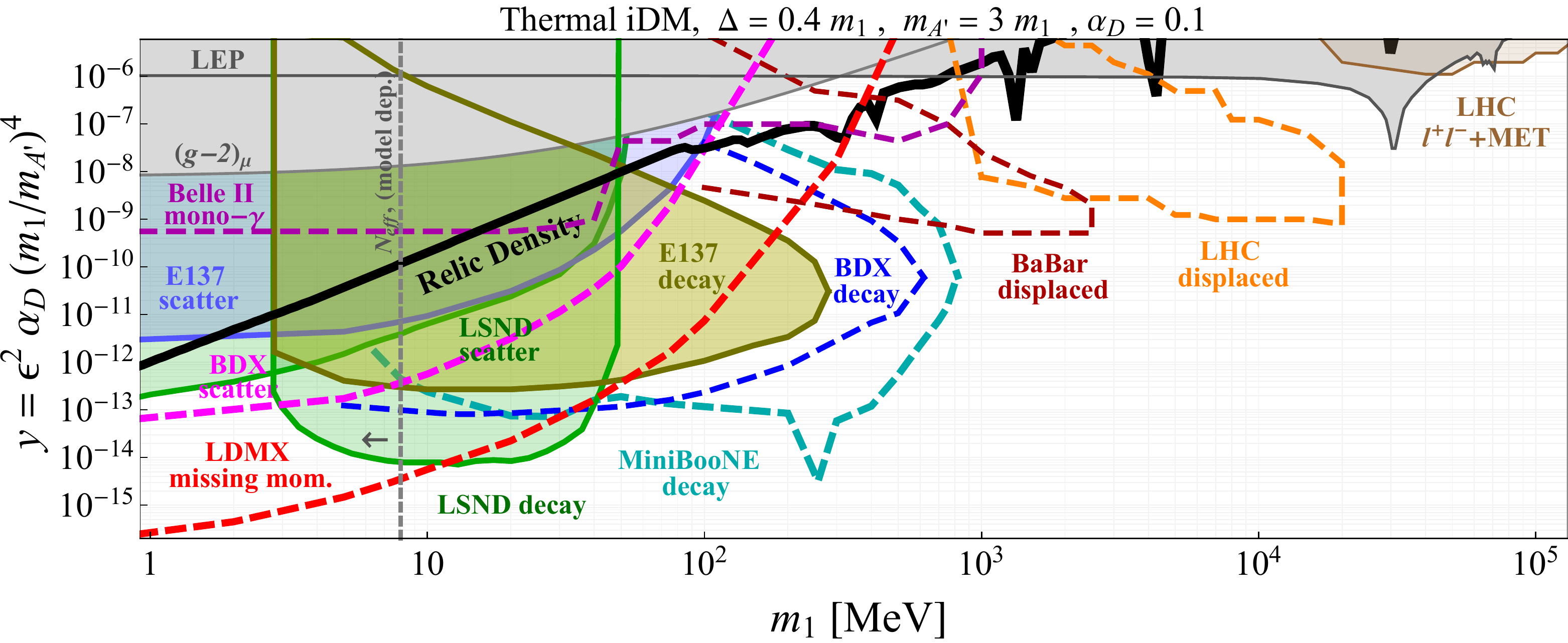}  ~~
  \caption{Same parameter space as the top-left and bottom-right panels of Fig.~\ref{fig:money}, but with the mass range extended out to the electroweak scale. Here we combine the results of this paper with the LHC, BaBar, and Belle II constraints and projections presented in Ref.~\cite{Izaguirre:2015zva}. The combined reach from the 
  sum of these efforts  suffices to cover nearly all remaining parameter space for thermal coannihilation; thermal DM models with masses below the MeV scale suffer 
  generic conflicts with $N_{\rm eff} $ \cite{Green:2017ybv} and masses above $\sim$ 100 TeV generically violate perturbative unitarity \cite{Griest:1989wd}. The only gaps
  not covered by this program of searches occur at very small mass splittings $\Delta \ll 0.01 m_1$, depicted in the lower left panel of Fig.~\ref{fig:varyalphaD}. For such small splittings, the decay searches become weak on account of the $\Gamma_{\chi_2} \propto \Delta^5$ scaling, and are not even kinematically allowed at low masses since $\Delta < 2m_e$.} 
   \label{fig:bigmoney}
\vspace{0cm}mon
\end{figure*}
  
We now have all the ingredients in place to assess the potential sensitivity of currently proposed fixed-target experiments to discover inelastic DM models. We begin with a brief discussion of existing constraints. 

The parameter space of inelastically coupled DM for mass scales beneath $\sim$ GeV is constrained by precision measurements, B-factories, and previous fixed-target experiments. On the precision front, the anomalous magnetic moment of the muon and electron constrains the interaction strength between the dark photon $\apr$ and the SM particles \cite{Pospelov:2008zw}. On the collider front, both LEP and BaBar set a bound for larger values of $\epsilon$. The former arises from the shift in $m_Z$ induced from mixing with $\apr$ \cite{Hook:2010tw}, and the latter from a monophoton and missing mass re-analysis \cite{Izaguirre:2013uxa, Essig:2013vha, Lees:2017lec}. 

Some of the strongest constraints from elastic DM arise from E137 \cite{Bjorken:1988as,Batell:2014mga}, an electron beam dump experiment, and LSND \cite{Auerbach:2001wg,deNiverville:2011it}, a proton beam fixed-target neutrino production experiment. Here, we reinterpret the constraints in terms of coannihilating DM. As discussed in Sec.~\ref{sec:generic} above, there are two qualitatively different signals: a scattering signal, where $\chi_{1,2}$ up- or downscatter at the detector and produce a recoiling target and possibly an $e^+e^-$ pair, and a decay signal, where $\chi_2$ survives to the detector and decays inside, producing an $e^+e^-$ pair. The reach of these experiments depends on their ability to distinguish these multiple signals. While E137 is only sensitive to total energy deposits, the angular resolution of LSND means it is potentially sensitive to well-separated $e^+/e^-$ pairs, which can be distinguished from the fake elastic events we used in estimating the sensitivity in this work and could enhance the sensitivity. However, this would require access to the LSND data as this signal of two charged tracks in the detector is not present in any published analysis. These existing constraints are illustrated in Fig.~\ref{fig:money} for $\alpha_D$ = 0.1, $m_{\apr} = 3m_1$, and various values of $\Delta$. For all but the smallest splittings, the combination of LSND and E137 covers a large portion of the thermal target in the 1-100 MeV range. However, for $2m_1+\Delta > m_{\pi^0}$, DM production through pions is kinematically forbidden, so we see sharp kinematic cutoffs at the pion threshold.

For comparison, Fig.~\ref{fig:varyalphaD} shows the effect of varying our benchmark parameters (each panel varies one detail relative to the top-left panel of Fig.~\ref{fig:money}) to demonstrate that these benchmarks are conservative and representative of the viable parameter space. 
In particular, the top row of Fig.~\ref{fig:varyalphaD} shows how the parameter space in the top-left plot of Fig.~\ref{fig:money} changes 
as we increase and decrease $\alpha_D$ while holding all other parameters fixed. Although there is slightly more viable parameter space for the large value 
$\alpha_D =0.5$, this choice is close to the perturbativity limit for abelian dark sectors \cite{Davoudiasl:2015hxa}, so we regard our benchmark choice as a 
representative and conservative value; choosing a smaller coupling excludes {\it more} parameter space on the $y$ vs. $m_1$ plane, as we see for $\alpha_D = \alpha$ in the same figure. In the bottom row of Fig.~\ref{fig:varyalphaD}, we vary our $\Delta$ and $m_{\apr}$ benchmarks from Fig.~\ref{fig:money}. In the bottom-left plot, we show the nearly-elastic case of $\Delta = 0.01m_1$, where the decay signal shuts off and the constraints are dominated by scattering. For comparison, we also show the recent MiniBooNE elastic scattering results \cite{Aguilar-Arevalo:2017mqx}, for which the beam energy is sufficiently large that the small 1\% mass splitting does not affect the reach. 

In the bottom-right plot, we show results for a larger hierarchy, $m_{\apr}/m_1=10$. For a given $m_1$, $\Delta$, and $\alpha_D$, the production rate is decreased as that event now arises from a much heavier $\apr$. If we parameterize the production rates at $m_{\apr}/m_1=3$ and $m_{\apr}/m_1=10$ as $N_3 \epsilon^2$ and $N_{10} \epsilon^2$, respectively, the total decay or scattering yield scales as $N_{3,10} \epsilon^4/m_{\apr}^4$. Thus, for a fixed event yield, $\epsilon$ scales linearly with $m_{\apr}$ but only as $N_{3,10}^{1/4}$. Far from any kinematic boundaries, the sensitivity in  $y \propto \epsilon^2 /m_{\apr}^4$ improves relative to the thermal target since the scaling with $m_{\apr}$ dominates the scaling with $(N_{3}/N_{10})^{1/4}$. However, the reach at large masses degrades as the $A'$ mass approaches the maximum available energy more rapidly and $A'$ production shuts off.

We now turn to the potential of new proposals to largely cover the entire parameter space motivated by thermal inelastic DM. We focus on three experiments representative of the setups we have previously discussed: MiniBooNE, BDX, and LDMX, which are proton beam dump, electron beam dump, and missing energy experiments, respectively. For comparison, we have also estimated the Belle II \cite{TheBelle:2015mwa} sensitivity only in the monophoton and missing mass channel by rescaling the BaBar result by the appropriate luminosity factor. As discussed in Sec.~\ref{sec:generic-formalism}, the dominant signal at MiniBooNE is $\chi_2$ decay in the detector whenever it is kinematically allowed. Since MiniBooNE has particle ID \cite{2005NIMPA.555..370Y,Patterson:2009ki}, electrons can in principle be distinguished from photons, and thus a well-separated $e^+/e^-$ pair and no other activity in the detector is a signal with few irreducible backgrounds. This stands in sharp contrast to the case of elastic DM scattering at MiniBooNE \cite{Aguilar-Arevalo:2017mqx}, which must always contend with an irreducible neutrino background. Note that the lower boundary of the decay curve is set by the energy threshold and angular resolution according to Eq.~(\ref{eq:deltamin}). We did not attempt a detailed simulation of the $\rho$ resonance, and thus the reach in $y$ of the region very close to $m_{A'} \sim m_\rho$ may differ slightly from what we show. The missing energy signal at LDMX dominates at low masses, while the decay reach for BDX is similar to that of MiniBooNE at high masses. Indeed, while the $A'$ production rate at proton beam experiments is a few orders of magnitude larger than at electron beam experiments, the higher luminosity and beam energy at BDX compensate to give a roughly similar reach. This is advantageous given that different models for the mediator could enhance or suppress production at proton beams compared to electron beams, so the combination of both experiments probes a wide range of models.

Fig.~\ref{fig:bigmoney} combines the results in this work with the results of Ref.~\cite{Izaguirre:2015zva} to show the thermal target over a wide range of DM masses. We see that except for a few isolated masses, the thermal target for coannihilating DM could be well-covered by all three planned experiments below $\sim$1 GeV, and by collider experiments from $\sim$1 GeV to 1 TeV. The scattering signals dominate at low mass below the kinematic threshold $\Delta = 2m_e$, while the decay signals dominate when kinematically allowed.

\section{Conclusion}
\label{sec:conclusion}
In this paper we have studied the fixed-target phenomenology of thermal dark matter with inelastic couplings to the SM and proposed
 a series of new searches for these interactions.  
These models are an instance of the general case where the relic abundance arises from thermal coannihilation between the halo DM candidate $\chi_1$ and an unstable excited state, $\chi_2$.
Since the heavier state decays away in the early universe, there is no annihilation at later times, and therefore no indirect detection. 
Furthermore, if the mass difference between these states exceeds $\sim 100 $ keV, upscattering at direct detection experiments is kinematically forbidden and loop-induced elastic scattering is small, so this scenario can likely only be discovered or falsified using accelerators. We leave the possibility of one-loop elastic scattering at recently proposed electron direct detection experiments for a future study.

At fixed-target experiments, the inelastic interaction responsible for setting the relic abundance yields a variety of observable signatures arising from the  
boosted $\chi_1 \chi_2$ system, which is produced in a proton or electron beam collision with target nuclei. Once produced, 
either state can scatter off particles in a downstream detector, thereby  generating an observable signal. In addition, the boosted $\chi_2$ can also
survive out to the detector and decay semi-visibly via $\chi_2 \to\chi_1 e^+e^-$ to directly deposit a visible signal as it passes through the 
active volume. 

Using these signatures, we have extracted existing constraints on this scenario by reinterpreting 
old LSND and E137 data. To this end, we have generalized the analyses in \cite{deNiverville:2011it} (for LSND) and \cite{Batell:2014mga} (for E137), which
focused on the scattering signatures of elastically coupled DM. In our 
analysis, we have demonstrated that there are several new signatures to which these older experiments are sensitive if DM couples
inelastically. In particular, we find that E137 and LSND already place nontrivial bounds on the parameter space
that yields sub-GeV thermal coannihilation for a variety of DM masses, mass splittings, and coupling strengths. 

We have also studied the prospects for future decay and scattering searches at the existing MiniBooNE (proton beam) experiment 
and the proposed BDX and LDMX (electron beam) experiments.
 We find that the combined reach of all scattering and decay searches at these experiments can comprehensively 
 test nearly all remaining parameter space,  thereby providing strong motivation for these efforts.  

This paper also extends earlier work  \cite{Izaguirre:2015zva}, which studied the collider phenomenology of inelastic thermal coannihilation 
models over the GeV -- TeV mass range. Our work complements this effort by working out the constraints and projections for the MeV--GeV range, thereby
providing a roadmap for covering thermal coannihilation over nearly all masses for which a thermal origin is viable (lower masses are in
conflict with early universe cosmology and higher masses generically violate perturbative unitarity in most models).  This full coverage, spanning
the results of both papers, is presented in Fig.~\ref{fig:bigmoney}.

Finally we note that other existing and future experiments may also have powerful reach to this class of models.
In particular, the proton beam experiments DUNE \cite{Acciarri:2016crz}, SeaQuest \cite{Nakahara:2011zz} (see forthcoming work by \cite{Asher}), SHiP \cite{Alekhin:2015byh},  and T2K \cite{Hayato:2005hq} all involve beam energies
in excess of 100 GeV, which can produce far more boosted DM than the beams considered in this 
work (all $<$ 10 GeV and below). Higher energies at these experiments can open up new production modes for the DM candidates (e.g. deep inelastic
scattering) and impart greater boosts to the DM system, which can profoundly affect the sensitivity projections for these setups. In addition, liquid argon detectors such as ICARUS \cite{Antonello:2013ypa} may be more optimized for seeing the two tracks characteristic of the decay signal.\footnote{We thank Maxim Pospelov for pointing this out.} However, 
working out the implications of these features is beyond the scope of this paper. 

Collider experiments may also probe the low-mass scenario we consider in this paper. The superior estimated reach of LHCb to visibly-decaying dark photons \cite{Ilten:2015hya, Ilten:2016tkc} suggests that LHCb will also have sensitivity to inelastic DM, especially when $\chi_2$ undergoes displaced decays within the detector. However, this requires a dedicated analysis, as the pairs of leptons in this case do \emph{not} reconstruct a resonance due to the invisible $\chi_1$. We leave this interesting analysis to future work.\footnote{We thank the anonymous referee for suggesting this analysis.} 

In addition, BaBar could be additionally sensitive to this topology through dedicated searches for monophoton and displaced leptons, as proposed by Ref.~\cite{Izaguirre:2015zva}. Similarly, through analogous searches, the larger luminosities at Belle II could provide unprecedented sensitivity to both displaced and long-lived decays. Parts of the parameter space with larger mass splittings can also lead to displaced vertices from $\chi_2$ decay, but requires a dedicated analysis for Belle II using displaced leptons; we defer this for a future study.

We look forward to the results of the numerous planned experiments on the horizon, and encourage them to pursue the inelastic DM implications of the signals we have discussed in this work.

\begin{acknowledgments}
{\it Acknowledgments:}  
We thank Brian Batell, Asher Berlin, Nikita Blinov, Maxim Pospelov, Philip Schuster, Brian Shuve, Tim Tait, Natalia Toro, Richard Van De Water, and Yiming Zhong for helpful conversations. We thank Jordan Smolinsky for pointing out typos in an earlier version of this paper. Fermilab is operated by Fermi Research Alliance, LLC, under Contract No. DE-AC02-07CH11359 with the US Department of Energy.   EI is supported by the United
States Department of Energy under Grant Contract desc0012704.
\end{acknowledgments}
 \medskip

\bibliography{LSNDiDMBib.bib}

\appendix

\section{Relic Abundance}
\label{sec:relic}

\begin{figure}[t] 
\hspace{-0.3cm}
\includegraphics[width=8.2 cm]{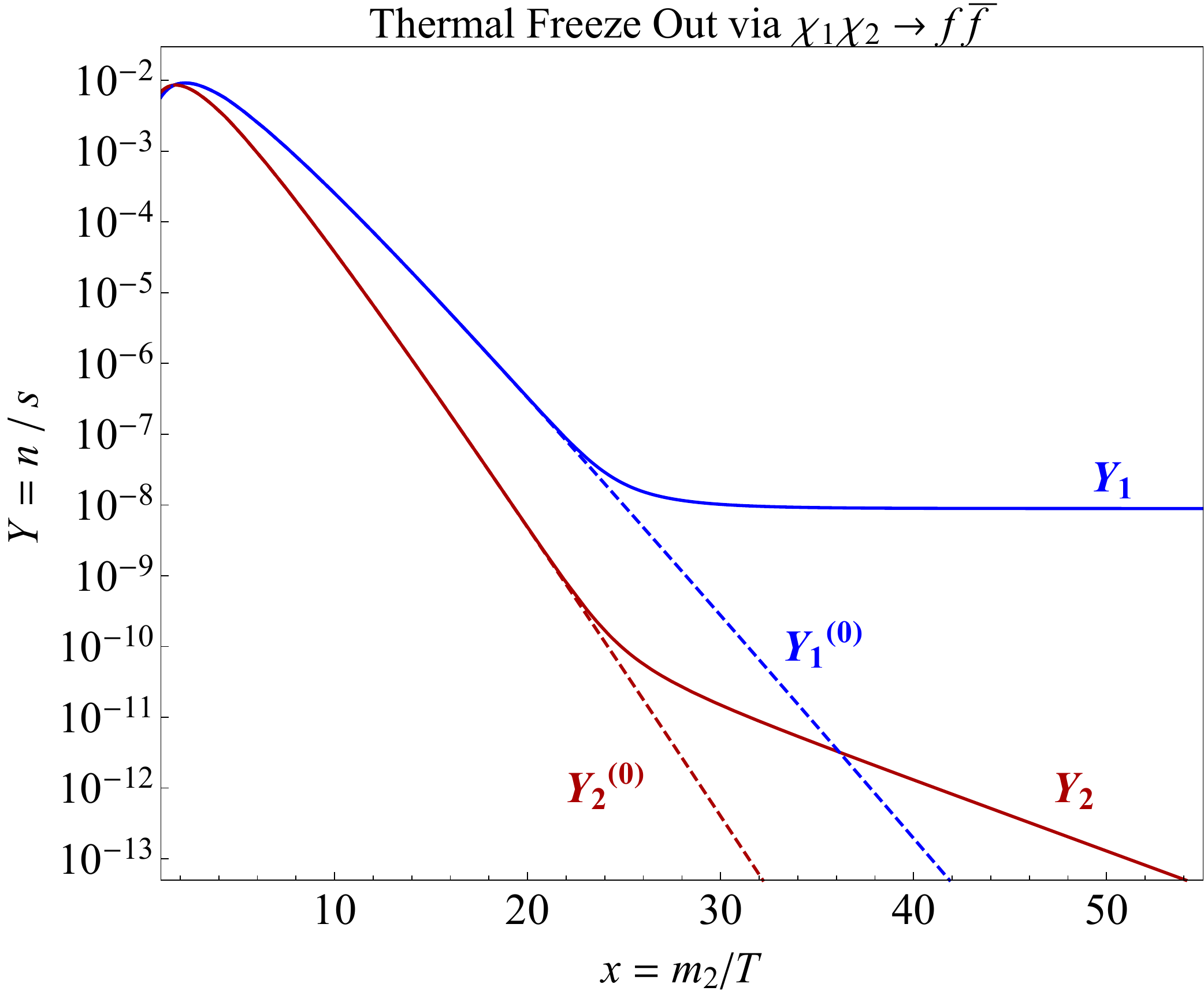}  ~~
  \caption{ 
   Example solution to the Boltzmann system in Eq. (\ref{eq:main-bolts}) for $m_1 = 50$ MeV, $\Delta = 0.3 m_1$,  $m_{A^\prime} = 3 m_1$, and $\langle \sigma v\rangle =  1.7 \times 10^{-23} \cm^3 \s^{-1}$, corresponding to $\Omega_{\chi_1} h^2 = 0.112$ at late times.}
   \label{fig:freezeout}
\vspace{0cm}
\end{figure}

The relic abundance of $\chi_1$ is governed by a Boltzmann equation whose collision terms involve thermally averaged coannihilation, decay, and inelastic scattering processes. Defining the dimensionless comoving yield as $Y_i \equiv n_i / s$, where $s(T) = 2 \pi^2 g_{s,*}T^3/45$ is the entropy density and $g_{*,s}$ is the number of entropic degrees of freedom, the Boltzmann system can be written as
\be\label{eq:main-bolts}
\frac{dY_{1,2}}{dx}  &=& - \frac{\lambda_{A}}{x^2} \Bigl(  Y_1 Y_2 -     Y^{(0)}_1 Y^{(0)}_ 2\Bigr)\pm x   \lambda_{D}    \Bigl( Y_2  -     \frac{Y_2^{(0)}}{Y_1^{(0)}} Y_1  \Bigr)
 \nonumber  \\
  &&~~~~~~~~~~~\pm \frac{\lambda_{S}}{x^2} Y^{(0)}_f   \Bigl(    Y_2  -   \frac{Y_2^{(0)}}{Y_1^{(0)}} Y_1   \Bigr)  ,
\ee
where $x\equiv m_2/T$ is a dimensionless time variable and $Y_i^{(0)}$ is the comoving equilibrium yield for species $i$. We define $\lambda_{A, S, D}$, to be the dimensionless annihilation, scattering, and decay rates
\be 
\lambda_A &=&        \frac{s(m_2)}{H(m_2)}   \langle    \sigma v(\chi_1 \chi_2 \to f\bar f) \rangle   \\
\lambda_{S} &=&  \frac{s(m_2)}{H(m_2)}       \langle \sigma v(\chi_2 f \to \chi_1f) \rangle    \\
\lambda_D &=& \frac{  \langle \Gamma( \chi_2 \to \chi_1 f\bar f ) \rangle }{  H(m_2)}, \label{eq:botlz-lambdaA} 
\ee
 and we use the  Hubble rate $H(T) \simeq 1.66 \sqrt{g_*} T^2/m_{Pl}$ during radiation domination,  where $g_*$ is the number of relativistic degrees of freedom, and $m_{Pl} = 1.2 \times 10^{19}\, \GeV$ is the Planck mass.
Solving this system yields the asymptotic value $Y_{1}^\infty$ at freeze-out near $x \sim 20$, which determines the relic abundance 
\be
\Omega_{\rm \small DM} = \frac{\rho_{\chi_1}}{\rho_{\rm cr}} = \frac{     m_1 s_0 Y_{1}^\infty  }{\rho_{\rm cr}} ,
\ee 
where $\rho_{\rm cr}= 8.1 h^2 \times 10^{-47} \, \GeV^4 $ is the critical density and $s_0 \simeq 2969 \, \cm^{-3}$  is the present day CMB entropy. 
An example solution to this system for a representative model point is presented in Figure \ref{fig:freezeout}.

\section{Scattering and Annihilation Rates}

\subsection{Coannihilation Rate}
\label{app:annihilation}

The amplitude for this $\chi_1(p_1) \chi_2(p_2) \to f(k_1) \bar{f}(k_2)$ coannihilation is 
\be 
{\cal A} = \frac{i \epsilon e g_D}{s - m_{\apr}^2}  \bar u(p_1)  \gamma^\mu v(p_2) \, \bar u(k_1) \gamma_\mu v(k_2),
\ee
so squaring and averaging initial state spins gives
\be
\langle|\mathcal{A}|^2\rangle &=& \frac{2\epsilon^2e^2g_D^2}{(s-m_{A'}^2)^2}
\bigg [\left(t-m_1^2-m_f^2\right)\left(t-m_2^2-m_f^2\right)\nonumber \\
  && + \left(u-m_1^2-m_f^2\right)\left(u-m_2^2-m_f^2\right) \nonumber\\
  && +2m_1m_2\left(s-2m_f^2\right)+2m_f^2\left(s-m_1^2-m_2^2\right) \nonumber\\
  && +8m_1m_2m_f^2\bigg ]. 
\ee
The differential cross section for this process is 
\be
\frac{d\sigma}{d\Omega_*} =  \frac{|\vec k_*|}{|\vec p_*|} \frac{\langle |{\cal A}|^2 \rangle}{64 \pi^2 s } ,
\ee
where $|\vec p_*|$ and $|\vec k_*|$ are the initial and final momenta in the CM frame and 
\be
\frac{|\vec k_*|}{|\vec p_*|} =  \sqrt{  \frac{(s- 2m_f^2)^2 - 4m_f^4}{ (s- m_1^2 - m_2^2)^2 - 4m_1^2 m_2^2}  }~.
\ee
Integrating this result, the total cross section becomes
\be \label{eq:fullcross_mA}
\sigma(s) &=& \frac{|\vec{k}_*|}{|\vec{p}_*|}\frac{8\pi\epsilon^2\alpha\alpha_D}{s(s-m_{A'}^2)^2}
\biggl[\frac{1}{8}s^2+\frac{2}{3}|\vec{p}_*|^2|\vec{k}_*|^2\nonumber\\
  &&+\frac{1}{2}m_{\chi}^2(s-2m_f^2)+\frac{1}{2}m_f^2(s-2m_{\chi}^2)+m_{\chi}^2m_f^2 \biggr], \nonumber\\
\ee
where we have taken the elastic limit $m_1=m_2=m_{\chi}$ for illustrative purposes, but we retain the full inelastic mass dependence in our calculations.

Generalizing the derivation in \cite{Gondolo:1990dk} for coannihilation, the thermally averaged cross section is 
\be
\langle \sigma v \rangle = \frac{1}{ N }
\int_{s_0}^\infty ds \, \sigma(s) (s- s_0)\sqrt{s}  K_1\left( \frac{\sqrt{s}}{T}\right) ,
\ee
where $s_0 \equiv (m_1 + m_2)^2$, the thermal averaging factor is 
\be
N = 8 m_1^2 m_2^2 T K_2\left( \frac{m_1}{T}\right)  K_2\left( \frac{m_2}{T}\right),
\ee
and $K_{1,2}$ are modified Bessel functions of the first and second kinds.

Note that in our calculation of the relic density, we account for annihilation to hadrons (e.g. $\chi_1 \chi_2 \to {\apr}^* \pi^+\pi^-$) by following the procedure described in 
\cite{Izaguirre:2015zva} where the final state phase space is extracted from the measured
distribution $R(s) \equiv \sigma(e^+e^- \to {\rm hadrons})/
\sigma(e^+e^- \to {\rm \mu^+\mu^-})$.

\subsection{DM-SM Scattering Cross Section}
\label{sec:DMSMpointlike}
For pseudo-Dirac DM, with mass splitting $\Delta \equiv m_2 -m_1$, 
the matrix element for the process $\chi_1(p_1) T(p_2) \to \chi_2(k_1) T(k_2)$ is  given by 
\be
{\cal A} = \frac{\epsilon e g_D }{(t - m^2_{A^\prime})} [\bar u(k_2)\gamma_\mu u(p_2) ] [ \bar u(k_1)\gamma^\mu  u(p_1)]   \hspace{3mm}, 
\ee
so the squared, spin-average matrix element is 
\be
\langle |{\cal A}|^2 \rangle
 &=& \frac{128\pi^2 \epsilon^2 \alpha \alpha_D }{(t - m^2_{A^\prime})^2}  \biggl[    
(k_1\cdot k_2)(p_1\cdot p_2)+ (k_2\cdot p_1)(p_2\cdot k_1) \nonumber \\ &&  -  m_1 m_2 (k_2\cdot p_2) -  m_T^2 (p_1\cdot k_1) + 2 m_1 m_2 m_T^2  
\biggr] .~~~~~~~~
\ee
The differential scattering cross section in the CM frame is  
\be \label{eq:xsecCM}
\frac{d\sigma}{d\Omega^*} &=&
\frac{1}{2\pi}\frac{d\sigma}{d\cos\theta^*} = \frac{  \langle |{\cal A}|^2 \rangle}{  64 \pi^2 s  }    \frac{  |\vec k^*|  }{   \left|  \vec p^*  \right| },
\ee
where initial/final state momenta satisfy
 \be
  |\vec p^{\,*}|^2 &=&  \frac{ (s-m_T^2-m_1^2)^2 - 4 m_T^2 m_1^2}{4s}, \\  |\vec k^{\,*}|^2 &=&  \frac{ (s-m_T^2-m_2^2)^2 - 4 m_T^2 m_2^2}{4s}.
 \ee
In terms of  lab frame quantities with a stationary target $T$,
\be 
s &=& (p_1 + p_2)^2 = m_1^2 + m_T^2 + 2m_T E_{p_1}, \\
k_1\cdot p_1 & =& -\frac{1}{2} ( 2m_T^2 - m_1^2 -m_2^2 - 2m_T E_{k_2}   ) \nonumber \\  &=&  E^*_{p_1}  E^*_{k_1}  - |\vec p^*||\vec k^*| \cos \theta^*,
\ee
so we can change variables to obtain the differential  recoil distribution 
 \be
   d\cos \theta^* = \frac{m_T}{  |\vec p^*|  |\vec k^*| } dE_{T},
 \ee
where $E_{T} \equiv E_{k_2}$ is the targets recoil energy. Thus, we have
\be
\frac{d\sigma}{dE_{T}} = \frac{  m_{T} \langle |{ \cal A}|^2 \rangle}{  32 \pi  s  \left|  {\vec p}^*  \right|^{2} },
\ee
which serves as an input into all detector scattering calculations for both proton and electron beam dump experiments. 

\section{Matrix Elements for DM Production and Detection}
\label{app:matrix_elements}
\subsection{Meson Decay}
The matrix element for pseudoscalar meson decay $\mzero (p_1) \to \gamma(k_1)\chi_1(k_2)\chi_2(k_3)$ ($\mzero=\pizero,\eta$) is given by
\be
\mathcal{A}_{\mzero\to\gamma\chi_1\chi_2} &=&
\frac{-i\epsilon e^2g_D}{4\pi^2f_{\mzero}}
\frac{\left(g_{\nu\lambda}-\frac{q_{\nu}q_{\lambda}}{m_{A'}^2}\right)}{s-m_{A'}^2+im_{A'}\Gamma_{A'}} \times \nonumber \\
&&
 \epsilon^{\mu\nu\alpha\beta}{k_1}_{\alpha}q_{\beta}\epsilon^*_{\mu}(k_1)
\left[\bar{u}(k_2)\gamma^{\lambda}v(k_3)\right],
\ee
where $f_{\mzero}$ is the meson decay constant, $\Gamma_{A'}$ is the total $A'$ width, $q\equiv p_1-k_1=k_2+k_3$, and $s\equiv q^2$. The spin-averaged square of the matrix element is
\be
&&\langle\left|\mathcal{A}_{\mzero\to\gamma\chi_1\chi_2}\right|^2\rangle =
\frac{4\epsilon^2\alpha^2\alpha_D}{\pi f_{\mzero}^2}\frac{1}{(s-m_{A'}^2)^2+m_{A'}^2\Gamma_{A'}^2}\nonumber \\
&& \left[(m_{\mzero}^2-s)^2(s+2m_1m_2)
   -8s(k_1\cdot k_2)(k_1\cdot k_3)\right. \nonumber\\
  && \left.+ 2(m_{\mzero}^2-s)(m_2^2-m_1^2)(k_1\cdot (k_2 - k_3))\right].
\ee
If $m_1+m_2\ll m_{A'} \ll m_{\mzero}$ then we can reasonably make the narrow width approximation \cite{Kahn:2014sra} and  take
\be
\frac{1}{(s-m_{A'}^2)^2+m_{A'}^2\Gamma_{A'}^2} \to \frac{\pi}{m_{A'}\Gamma_{A'}}\delta(s-m_{A'}^2)
\ee

The decay width is given by
\be
d\Gamma_{\mzero\to\gamma\chi_1\chi_2} &=& \frac{1}{4\pi m_{\mzero}}\frac{\beta(m_{\mzero}^2,0,s)}{32\pi^2}\frac{\beta(s,m_1^2,m_2^2)}{32\pi^2} \times\nonumber \\
&&\langle\left|\mathcal{A}_{\mzero\to\gamma\chi_1\chi_2}\right|^2\rangle d\Omega_\gamma^* d\Omega_{\chi}^*ds,
\ee
where we have defined the function
\be
\label{eq:beta}
\beta(s_{12},s_1,s_2) = \sqrt{1 - \frac{2(s_1+s_2)}{s_{12}} + \frac{(s_1-s_2)^2}{s_{12}^2}}.
\ee
Here, $d\Omega_{\gamma}^*$ refers to angles in the $\mzero$ rest frame and $d\Omega_{\chi}^*$ refers to angles in the $\chi_1\chi_2$ CM frame. 

\subsection{Excited State Decay}
The matrix element for $\chi_2(p_1)\to\chi_1(k_1)f(k_2)\bar{f}(k_3)$ is given by
\be
\mathcal{A}_{\chi_2\to\chi_1f\bar{f}} &=&
\frac{\epsilon e g_D}{(s-m_{A'}^2) + im_{A'}\Gamma_{A'}}\left(g_{\mu\nu}-\frac{q_{\mu}q_{\nu}}{m_{A'}^2}\right) \nonumber \\
&& \times \left[\bar{u}(k_1)\gamma^{\mu}u(p_1)\right]\left[\bar{u}(k_2)\gamma^{\nu}v(k_3)\right],
\ee
where again $\Gamma_{A'}$ is the total $A'$ decay rate,  $q\equiv p_1-k_1$, and $s\equiv q^2$. In this paper we will consider $f = e^-$ exclusively; decay to muons is allowed only for the largest masses and splittings but may provide a distinctive signal at higher-energy experiments.
The spin-averaged square is
\be
&&\langle\left|\mathcal{A}_{\chi_2\to\chi_1f\bar{f}}\right|^2\rangle = \frac{16\epsilon^2 e^2g_D^2}{(s-m_{A'}^2)^2+m_{A'}^2\Gamma_{A'}^2} \nonumber \\
&&\times \left[(k_2\cdot k_1)(k_3\cdot p_1)
  + (k_2\cdot p_1)(k_3\cdot k_1) \right.\nonumber \\
 && \left.  + m_f^2(k_1\cdot p_1)
  -m_1m_2(k_2\cdot k_3)
  - 2m_1m_2m_f^2\right].
\ee
We note that because we only consider $m_{A'}>m_1$ and $\Delta<m_1$ in this paper, the $A'$ is always off-shell and we never make the narrow width approximation for $\chi_2$ decays.

The decay width is given by
\be
d\Gamma_{\chi_2\to\chi_1f\bar{f}} &=& \frac{1}{4\pi m_2}\frac{\beta(m_2^2,m_1^2,s)}{32\pi^2}\frac{\beta(s,m_f^2,m_f^2)}{32\pi^2} \nonumber \\
&& \times\langle\left|\mathcal{A}_{\chi_2\to\chi_1f\bar{f}}\right|^2\rangle d\Omega_1^* d\Omega_f^*ds,
\ee
where $\beta$ is defined as in \Eq{eq:beta}, $d\Omega_1^*$ refers to angles in the $\chi_1$ rest frame, and $d\Omega_f^*$ refers to angles in the $f\bar{f}$ CM frame.

\subsection{DM-SM Scattering}
\label{sec:DMSMFF}
The tree-level matrix elements for scattering $\chi_i(p_1) T(p_2) \to \chi_j(k_1) T(k_2)$ off a pointlike fermionic target have already been computed above in \App{sec:DMSMpointlike}. Since we will also be interested in scattering off targets with substructure such as nucleons and nuclei, we consider the more general scattering process where $T$ is a fermionic target with both a monopole and dipole coupling to electromagnetism.
The matrix element for this process is given by
\be
\mathcal{A}_{\chi_iT\to\chi_jT} &=& \frac{\epsilon e g_D}{t-m_{A'}^2}\left(g_{\mu\nu}-\frac{q_{\mu}q_{\nu}}{m_{A'}^2}\right) \nonumber\\
&& \times\left[\bar{u}(k_1)\gamma^{\mu}u(p_1)\right]\left[\bar{u}(k_2)\Gamma^{\nu}u(p_2)\right],
\ee
where $q\equiv p_1-k_1$ is the four-momentum carried by the virtual photon and $t\equiv q^2$ is the Mandelstam variable.
The Lorentz structure $\Gamma^{\mu}$ at the target vertex is
\be
\Gamma^{\mu} = F_1(q^2)\gamma^{\mu} + F_2(q^2)\frac{iq_{\nu}\sigma^{\mu\nu}}{2m_f}.
\ee
Here, $\sigma^{\mu\nu}\equiv\frac{i}{2}\left[\gamma^{\mu},\gamma^{\nu}\right]$, and $F_1(q^2)$ and $F_2(q^2)$ are the electric monopole and dipole form factors which depend on the target $T$.
For the purposes of this paper, we take
\be
F_1 = \left\{\begin{matrix}
\frac{1}{(1-q^2/m_p^2)^2} & T=p\\
0 & T=n\\
Z & T=N\\
\end{matrix}\right.
\ee
and
\be
F_2 = \left\{\begin{matrix}
\frac{\kappa_p}{(1-q^2/m_p^2)^2} & T=p\\
\frac{\kappa_n}{(1-q^2/m_n^2)^2} & T=n\\
0 & T=N\\
\end{matrix}\right.
\ee
with $\kappa_p\approx1.79$ and $\kappa_n\approx-1.9$ \cite{deNiverville:2011it}.

The spin-averaged square of the matrix element is
\be
&&\langle\left|\mathcal{A}_{\chi_iT\to\chi_jT}\right|^2\rangle = \frac{16\pi^2\epsilon^2\alpha\alpha_D}{(t-m_{A'}^2)^2} \times  \\
&& \Big\{ (F_1+F_2)^2\left[t(p_1+k_1)^2 + 3t(t-\Delta^2)-(m_2^2-m_1^2)^2\right] \nonumber\\
&& \left. +\left(F_1^2+\tau F_2^2\right)\left[\left((p_1+k_1)\cdot(p_2+k_2)\right)^2+(k_2+p_2)^2(t-\Delta^2)\right]\right\}, \nonumber
\ee
where $\tau\equiv-\frac{t}{4m_T^2}$.

For an incoming $\chi_i$ with the target at rest in the lab frame, the lab-frame differential cross section is then given by
\be
\frac{d\sigma_{\chi_iT\to\chi_jT}}{d\Omega^*} &=& \frac{1}{2E_{\chi}}\frac{1}{2m_T}\frac{1}{|\vec{v}_{\chi}|}\frac{\beta(s,m_j^2,m_T^2)}{32\pi^2} \nonumber\\
&&\times\langle\left|\mathcal{A}_{\chi_iT\to\chi_jT}\right|^2\rangle,
\ee
where $\beta$ is defined as in \Eq{eq:beta}.

We note that for the case of coherent scattering off of nuclei ($T=N$), we make the additional insertion of the Helm form factor \cite{Izaguirre:2014dua,Duda:2006uk}.

\section{Monte Carlo Techniques}
\label{app:monte_carlo}
Our simulation performs two distinct operations: production of $\chi_1$ and $\chi_2$ pairs and the detection of $\chi_1$ $\chi_2$ pairs in a detector.
Many of the generic techniques used in our simulations such as numerical phase space integration, rejection sampling of differential probability distributions, and computations utilizing detector geometries were borrowed from or influenced by \verb!BdNMC! \cite{deNiverville:2016rqh}.

\subsection{DM Production Simulation for Proton and Electron Beams}
We use two simulation pipelines, one for proton beams and one for electron beams. Our proton beam production simulation takes as input an unweighted list of four-momenta from one of the DM progenitors that we considered: $\pi^0$, $\eta$, and $A'$ from dark bremsstrahlung.
The output of the production simulation is an unweighted list of $\chi_1$ and $\chi_2$ particles produced.
Each $\chi_1 \chi_2$ four-momentum pair in the output list is randomly sampled from the differential decay rate of the progenitor via a rejection sampling method similar to that used in \verb!BdNMC!. 

For electron beam production, we used a modified version of \texttt{Madgraph@NLO} \cite{Alwall:2014hca} from Ref.~\cite{Chen:2017awl}, which creates a new physics model which contains new particle containers for the nucleus, $\apr$ and $\chi_{1,2}$ system. It utilizes elastic and inelastic atomic form factors from Ref.~\cite{Kim:1973he}.
The elastic component is given by 
\be
G_{2,el}(t)= \left(\f{a^2 t}{1+a^2 t} \right)^2
\left(\f{1}{1+t/d} \right)^2 Z^2,
\ee
where $a=111\,Z^{-1/3}/m_e$, and $d=0.164 \mbox{ GeV}^2 A^{-2/3}$.  
We also include a quasi-elastic (inelastic) term:
\be
G_{2, in}(t)= \left(\f{a'^2 t}{1+a'^2 t} \right)^2 \left(\f{1+\f{t}{4
    m_p^2} (\mu_p^2-1)}{(1+\f{t}{0.71\,{\rm GeV}^2})^4} \right)^2 Z,
\ee
where $a'=773 \,Z^{-2/3}/m_e$, $m_p$ is the proton 
mass, and $\mu_p=2.79$.  
The general form factor is then
\be
\Phi \equiv \int_{t_{min}}^{t_{max}}  dt \f{t-t_{min}}{t^2} (G_{2,el}(t)+G_{2,in}(t)). \label{ChiExp}
\ee

\subsection{DM Detection Simulation for Neutrino Detectors}
\label{app:DMdetection}
Our detector simulation takes as input the list of $\chi_1 \chi_2$ four-momentum pairs from the production simulation.
We assume that the various production processes all take place at the beam stop, so that the trajectories of each DM particle is well defined.
There are three ways that a $\chi_1 \chi_2$ pair produced at the beam stop can be detected:
the $\chi_2$ can reach the detector and either decay or downscatter or the $\chi_1$ can reach the detector and upscatter.

For each $\chi_2$ in the input list, a decay length $x$ is randomly selected from the appropriate exponential distribution.
If the $\chi_2$ trajectory intersects with the surface of the detector, we calculate the path length $l$ along the trajectory from the beam stop to the detector.
If $x>l$, i.e. if the $\chi_2$ persists until it reaches the detector, then the $\chi_2$ can be detected via direct decay or downscattering.
If $x<l$ or if the $\chi_2$ does not intersect the detector, we decay the $\chi_2\to\chi_1\epm$ via rejection sampling of the differential decay rate.
We then process the resulting $\chi_1$ four-momentum in exactly the same way as a $\chi_1$ produced from primary progenitors, but now accounting for the fact that this $\chi_1$ trajectory starts at the point where the $\chi_2$ decayed rather than the beam stop.

Each $\chi_2$ that reaches the detector can either decay or downscatter.
We process decay events by weighting by the total probability for the $\chi_2$ to decay inside the detector and performing the decay $\chi_2\to\chi_1\epm$ via rejection sampling of the differential decay rate.
We then process the final state $\epm$ pair by applying the kinematic cuts described in \Sec{sec:bounds-projections}.
We process scattering events by summing over targets and consider scattering off of each target $T$ independently.
To avoid double counting with decay events, we make the conservative assumption that the $\chi_2$ can only scatter if it does not decay in the detector.
We therefore weight by the probability that the $\chi_2$ persists through the detector and the independent probability that the $\chi_2$ scatters in the detector and perform the scattering $\chi_2 T \to \chi_1 T$ by sampling from a uniform distribution in the center of mass variables $\cos\theta$ and $\phi$, which uniquely specify the final state kinematics.
Because we sample our final state kinematic variables from a uniform distribution rather than the true distribution of final states,
\begin{equation}
\mathcal{P}(\cos\theta,\phi) = \frac{1}{\sigma_T}\frac{d\sigma_T}{d\Omega},
\end{equation}
which is proportional to the differential cross section, we must also weight each event by the factor $4\pi\mathcal{P}(x_1,...,x_n)$. 
This weighting scheme enables a cancellation of the total cross section $\sigma_T$, making a Monte Carlo computation of $\sigma_T$ unnecessary and thereby saving significant computational complexity.
The cancellation occurs because each event is also weighted by the total probability for scattering which is $\propto \sigma_T$ when Taylor-expanded.
Once the scattering final state is sampled and the weights computed, we apply the cuts described in \Sec{sec:bounds-projections} to the recoiling target.

For each $\chi_1$ that reaches the detector, we process the scattering $\chi_1 T \to \chi_2 T$ using the same method as $\chi_2$ downscattering.
Additionally, we weight by the probability that the upscattered $\chi_2$ will decay in the detector and perform the decay via rejection sampling.
The $\epm$ pair from the de-excitation and the recoiling target are both included when we apply the kinematic cuts described in \Sec{sec:bounds-projections}.

\end{document}